\documentclass[twocolumn, tighten, twocolappendix]{aastex631}

\newcommand{\rev}[1]{{#1}} 
\newcommand{\revv}[1]{{#1}} 
\graphicspath{{figs/}}
\usepackage{comment}

\begin{document}

\title{Mapping the merging zone of late infall in the AB Aur planet-forming system}

\correspondingauthor{J. Speedie}
\email{jspeedie@uvic.ca}

\author[0000-0003-3430-3889]{Jessica Speedie}
\affiliation{Department of Physics \& Astronomy, University of Victoria, Victoria, BC, V8P 5C2, Canada}

\author[0000-0001-9290-7846]{Ruobing Dong} 
\affiliation{Department of Physics \& Astronomy, University of Victoria, Victoria, BC, V8P 5C2, Canada}
\affiliation{Kavli Institute for Astronomy and Astrophysics, Peking University, Beijing 100871, People’s Republic of China; rbdong@pku.edu.cn}

\author[0000-0003-1534-5186]{Richard Teague} 
\affiliation{Department of Earth, Atmospheric, and Planetary Sciences, Massachusetts Institute of Technology, Cambridge, MA 02139, USA}

\author[0000-0003-3172-6763]{Dominique Segura-Cox} 
\affiliation{Department of Astronomy, The University of Texas at Austin, 2500 Speedway, Austin, TX 78712, USA}
\affiliation{Department of Physics and Astronomy, University of Rochester, Rochester, NY 14627-0171, US}

\author[0000-0002-3972-1978]{Jaime E. Pineda} 
\affiliation{Max Planck Institute for Extraterrestrial Physics, Gießenbachstraße 1, 85748, Garching bei München, Germany}

\author[0000-0001-7764-3627]{Josh Calcino} 
\affiliation{Department of Astronomy, Tsinghua University, 30 Shuangqing Rd, 100084 Beĳing, China}

\author[0000-0003-4663-0318]{Cristiano Longarini} 
\affiliation{Institute of Astronomy, University of Cambridge, Madingley Road, Cambridge, CB3 0HA, United Kingdom}

\author[0000-0002-8138-0425]{Cassandra Hall} 
\affiliation{Department of Physics and Astronomy, The University of Georgia, Athens, GA 30602, USA}
\affiliation{Center for Simulational Physics, The University of Georgia, Athens, GA 30602, USA}

\author[0000-0002-0675-276X]{Ya-Wen Tang} 
\affiliation{Academia Sinica, Institute of Astronomy and Astrophysics, 11F of AS/NTU Astronomy-Mathematics Building, No.1, Sec. 4, \\Roosevelt Rd., Taipei, Taiwan}

\author[0000-0002-3053-3575]{Jun Hashimoto} 
\affiliation{Astrobiology Center, National Institutes of Natural Sciences, 2-21-1 Osawa, Mitaka, Tokyo 181-8588, Japan}
\affiliation{Subaru Telescope, National Astronomical Observatory of Japan, Mitaka, \\Tokyo 181-8588, Japan}
\affiliation{Department of Astronomy, School of Science, Graduate University for Advanced Studies (SOKENDAI), \\Mitaka, Tokyo 181-8588, Japan}

\author[0000-0002-4044-8016]{Teresa Paneque-Carreño} 
\affiliation{Leiden Observatory, Leiden University, P.O. Box 9513, NL-2300 RA Leiden, the Netherlands}
\affiliation{European Southern Observatory, Karl-Schwarzschild-Str 2, 85748 Garching, Germany}

\author[0000-0002-2357-7692]{Giuseppe Lodato} 
\affiliation{Università degli Studi di Milano, Via Celoria 16, 20133, Milano, Italy}

\author[0000-0002-2611-7931]{Bennedetta Veronesi} 
\affiliation{Univ Lyon, Univ Lyon1, Ens de Lyon, CNRS, Centre de Recherche Astrophysique de Lyon UMR5574, F-69230, \\Saint-Genis,-Laval, France}

\begin{abstract}

Late infall events challenge the traditional view that planet formation occurs without external influence. 
Here we present deep ALMA $^{12}$CO $J=2-1$ and SO $J_{N}=5_6-4_5$ observations toward AB Aurigae, a Class II disk system with 
strong signs of gravitational instability and ongoing planet formation. 
By applying Keplerian and anti-Keplerian masks, we separate disk-like and non-disk-like motions of $^{12}$CO, considering the two outputs as the `disk' and `exo-disk' (out of disk) emission components, respectively.  
The disk component of $^{12}$CO extends to $\sim 1600$ au in radius and exhibits a stunningly rich architecture of global spiral structure. The exo-disk emission consists predominantly of three spiral structures --S1, S2 and S3-- whose projections are co-spatial with the disk. We successfully reproduce their 
trajectories 
with a ballistic accretion flow model, finding that S1 and S2 (both redshifted) are infalling toward the disk from in front, and S3 (blueshifted) is infalling from behind. Where the terminal ends of S1 and S2 become indistinguishable from the disk, we observe a brightness peak in SO emission $2.5\times$ the azimuthal average of a background SO ring.  
This merging zone lies within a relatively confined region $15\degr-100\degr$ east of north, and between $\sim150-300$ au from the star, 
at scales relevant to 
where planet candidates have been previously identified. The AB Aur system provides a unified picture of late infall inducing replenishment of the disk, triggering gravitational instability, and modifying the conditions of forming planets.

\end{abstract}

\keywords{Protoplanetary disks(1300) --- Planet formation(1241) --- Gravitational instability(668)}

\section{Introduction} \label{sec:introduction}

Recent scattered light and molecular observations are revealing more and more instances of large-scale narrow accretion filaments –`streamers'– delivering material to {circumstellar} 
disks, a process referred to as `late-stage infall' in the case of Class II disks \cite[e.g., ][]{ginski2021-SUAur, huang2021-MAPS-GMAur, garufi2022-hltau-dgtau, harada2023-DKCha, pineda2023-ppvii, huang2024-drtau-so, gupta2024-tipsy-hltau-scra}. Modeling studies have shown that infalling material can have radical consequences for the disk chemistry and structure, including localized shock heating, the formation of pressure bumps, {vortices or spiral arms}, 
and disk warps or misalignments \cite[e.g.,][]{hennebelle2017-infall-spirals, vangelder2021-shocks-sulphur, kuffmeier2021-infall-misalignments, kuznetsova2022-infall-pressurebumps-vortices, calcino2024-anatomyofafall}. 
Late infall in older Class II disks undergoing planet formation is thus highly relevant to the local conditions of forming planets.

AB Aurigae (AB Aur) is a 3.9 - 4.4 Myr old \citep{dewarf2003, beck2019, garufi2024-destinys-taurus} 
Class II {young stellar object} \cite[{YSO;}][]{henning1998, bouwman2000} at a distance of $155.9 \pm 0.9$ pc \citep{gaiaDR3-2023}. 
Several candidate protoplanets have been identified or predicted amongst the disk's spiral arms \citep{dong2016-spirals-scatteredlight, tang2017-abaur12COspirals, boccaletti2020, currie2022-abaurb, zhou2023-abaurb, biddle2024-pabeta-ABAur, currie2024-pabeta-ABAur}, suggesting planet formation is underway. 
Optical and {(sub)mm} observations of the local environment have led numerous studies to propose that the AB Aur system is currently experiencing infall from its surroundings \cite[][]{nakajima-golimowski1995-palomar, grady1999-hst, tang2012-abaur-envelope, dullemond2019-cloudlet, riviere2020-rosetta1, kuffmeier2020-late-encounters, gupta2023-reflection-nebulae}. The disk itself has long been suspected to be gravitationally unstable \citep{fukagawa2004}, and recently found to exhibit the predicted kinematic markers of gravitational instability \citep{hall2020, longarini2021} in observations of $^{13}$CO and C$^{18}$O emission 
{with the Atacama Large Millimeter/submillimeter Array} \cite[{ALMA;}][]{speedie2024}.  

In this letter, we use high spectral resolution ALMA observations of $^{12}$CO $J=2-1$ and SO $J_{N}=5_6-4_5$ emission to investigate the hypothesis that late infall is inciting {gravitational instability (GI)} in this relatively evolved system. 
Unlike other systems fed by single-streamers observed on spatial scales that dwarf their end destination \cite[e.g.,][]{pineda2020-natast, valdiviamena2022-peremb50, valdiviamena2023-b5, flores2023-edisk-irs63, cacciapuoti2024-dusty-shrimp, gupta2024-tipsy-hltau-scra}, the 
AB Aur system hosts \textit{multiple} `out-of-plane' spiral arms \textit{on disk scales}, detected in previous $^{12}$CO observations \citep{tang2012-abaur-envelope}. In the 2D sky plane, these structures appear co-spatial with the disk, necessitating new techniques for isolating their emission and presenting  a unique opportunity to study the interface between the disk and infalling material in detail.

\section{Observations} \label{sec:data}

We observed AB Aur with ALMA in Cycle 8 under program ID 2021.1.00690.S (PI: R. Dong). 
Here we present the $^{12}$CO $J=2-1$ {($\nu_{\rm rest}=230.538$ GHz, $E_{\rm up}=16.6$ K)} and SO $J_{N}= 5_6-4_5$ ({$\nu_{\rm rest}=219.949$ GHz}, $E_{\rm up}=35.0$ K) molecular emission line components from this program, which were reduced and imaged concurrently with the $^{13}$CO $J=2-1$ and C$^{18}$O $J=2-1$ components presented in \citet{speedie2024}. 
We refer the reader to that work for data processing details, and 
briefly reproduce the key aspects {below}\footnote{An online guide to the reduction and imaging for the full program is available at \url{https://jjspeedie.github.io/guide.2021.1.00690.S}}.

\subsection{{Calibration}} \label{subsec:calibration}

Measurements were taken in a short-baseline configuration C-3 (2 execution blocks) and a long-baseline configuration C-6 (6 execution blocks), using the Band 6 receivers \citep{ediss2004-band6}. We configured two spectral windows to target the $^{12}$CO $J=2-1$ and SO $J_{N}= 5_6-4_5$ lines, 
and dedicated a single {$2.0$ GHz} spectral window to sampling the continuum, centered at $233.012$ GHz. 
Using the continuum data, we aligned the eight execution blocks to a common phase center in the \textit{uv}-plane (employing the {\sc exoALMA} alignment script; {Loomis et al. submitted}). {Manual self-calibration was first performed on the short-baseline continuum data, starting with 6 rounds of phase-only calibration down to a solution interval of 18 s, followed by 1 round of amplitude+phase calibration on the interval of a scan length, resulting in a SNR increase by a factor of 2.2.  
Self-calibration subsequently continued on the concatenated short- and long-baseline continuum data, undergoing 4 rounds of phase-only calibration down to 60 s followed by 1 round of amplitude+phase calibration on the interval of a scan length, yielding a SNR increase by a factor of 1.5.}  Finally, the phase center shifts and self-calibration solutions acquired with the continuum data were applied to the line data, and continuum subtraction was done in the \textit{uv}-plane. 

\subsection{{Imaging}} \label{subsec:imaging}

\begin{table*}
\centering 
  \caption{  Molecular lines and properties of the imaged ALMA data cubes {(program ID 2021.1.00690.S).}} \vspace{0.5em}
  \label{tab:data}
  \begin{tabular}{ccccccc}
  \hline
    \textbf{Transition} & \textbf{Rest Frequency} & \textbf{Channel Width} & \texttt{robust} & \textbf{Beam}  & \textbf{rms}      & \textbf{JvM} $\epsilon$ \\
           & (GHz)            & (m/s)          &        & ($'' \times ''$, deg) & (mJy/beam) &             \\ \hline
$^{12}$CO $J=2-1$   & 230.538        & 42.0           & 0.5    & $0.23\times0.17$, 0.2           & 1.55     & 0.49        \\
SO $J_{N}=5_6-4_5$   & 219.949        & 84.0           & 1.5    & $0.39\times0.28$, 171.2         & 0.63     & 0.34      \\
  \hline
  \end{tabular}
\end{table*}

Imaging was undertaken with the {\sc CASA} \texttt{tclean} task. The $^{12}$CO data immediately presented two challenges: (i) {the presence of} large-scale diffuse emission with non-Keplerian morphologies, {for which generating the optimal CLEAN mask became a protracted and nebulous goal;} and (ii) the presence of negative bowling {\citep{braun-walterbos1985, holdaway1999}}, particularly in central channels.  
{Negative bowling, where the recovered (positive) emission appears to sit in a bowl of negative flux \cite[e.g. Fig. 7 of][]{faridani2018-shortspacingcorrection}, is a fundamental interferometer image artifact caused by the presence of} {emission on spatial scales larger than the reciprocal of the shortest measured baselines (known as the Maximum Recoverable Scale, MRS) 
and can only be solved by obtaining additional observations sampling the visibility function at smaller $uv$-spacings. 
{Animated $^{12}$CO channel maps illustrating the negative bowling are available (Figure \ref{appfig:environment}) in the online article.} 
The MRS} of our short-baseline configuration, C-3, is $\sim7\arcsec$ at $230$ GHz \citep{alma-technical-handbook2019}.  

We adopted the following strategies {in response to} both of the above two issues: First, we cleaned 
with a broad mask\footnote{We used the \texttt{tclean} arguments \texttt{usemask=`pb'} and \texttt{pbmask=0.2}, which sets a cleaning mask extending to where the 12m antenna primary beam gain reaches the $20\%$ level, which is usually considered the edge of the FOV.} encompassing all emission within the field of view (FOV), and accordingly, we cleaned conservatively (to a threshold of $5\times$ the rms noise measured in $20$ line-free channels of the dirty cube). 
 Second, we forced frequent major cycles\footnote{The $^{12}$CO cube with \texttt{robust=0.5} underwent 282 major cycles.} to repeatedly re-populate the $uv$-plane and interpolate into the missing short \textit{uv}-spacings. 
These strategies borrowed from the philosophy of {\sc PHANGS-ALMA} \citep{leroy2021-phangsalma}. 
We used the multiscale deconvolution algorithm \citep{cornwell2008} with 
Gaussian deconvolution scales [$0.02''$, $0.1''$, $0.3''$, $0.6''$, $1.0''$], with an additional largest scale of $2.0''$ appended for $^{12}$CO. 
We adopted a Briggs robust weighting scheme, and imaged in LSRK velocity channels at $42$ m/s for $^{12}$CO and $84$ m/s for SO.  
Unless otherwise specified, the images presented in this work use \texttt{robust=0.5} for $^{12}$CO and \texttt{robust=1.5} for SO, and are JvM corrected \citep{JvM1995-correction, czekala2021-maps2} and primary beam corrected. 
Table \ref{tab:data} lists details for each image cube.  
The rms noise is measured within an annulus of $4\arcsec$ inner radius and $8\arcsec$ outer radius over the first and last $5$ channels of the cube.

{We used the \texttt{bettermoments} package \citep{teague2018-bettermoments, teague2018-robust-linecentroids} to generate maps of peak intensity, velocity-integrated intensity (moment 0), and intensity-weighted velocity (moment 1) throughout this work.} 
We applied $\sigma$-clipping at $5\times$ the rms noise for $^{12}$CO and $3\times$ the rms noise for SO.

\subsection{{Overview with Standard Moment Maps}} \label{subsec:fiducial-maps}

\begin{figure*}
\centering
\includegraphics[width=\textwidth]{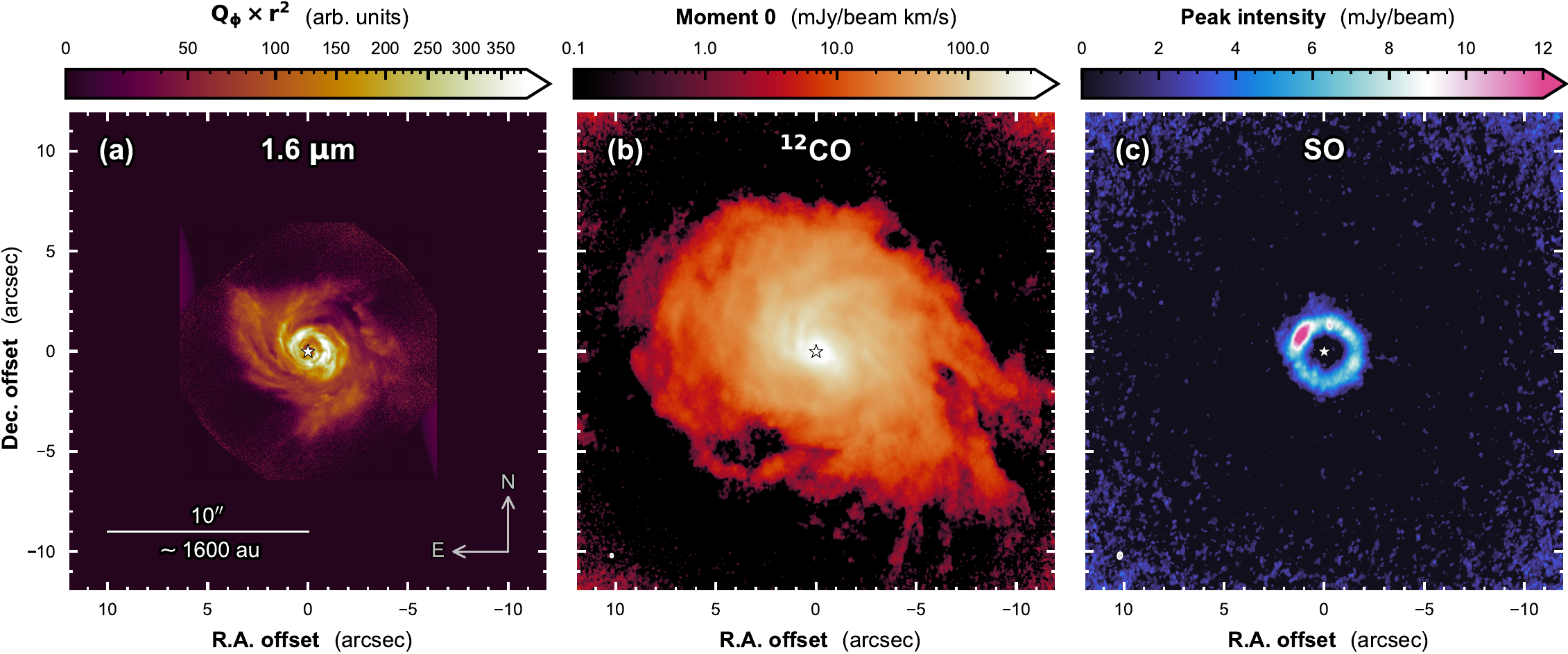}
\caption{\textbf{Introducing our ALMA observations.} \textbf{(a)} VLT/SPHERE \textit{H}-band scattered light image at 1.6 $\mu$m  \citep{boccaletti2020}, shown for scale.
\textbf{(b)} ALMA $^{12}$CO $J=2-1$ integrated intensity map, collapsed uniformly over line-of-sight velocities $\pm 6$ km/s about $v_{\rm sys}$. 
Emission extends $\sim 20''$ ($\sim 3200$ au) in diameter, roughly half the diameter of the 12m-antenna field of view in Band 6. \textbf{(c)} ALMA SO $J_{N}=5_6-4_5$ peak intensity map. The ring takes a brightness peak in the northeast. 
}
\label{fig:1}
\end{figure*}

Figure \ref{fig:1} introduces our ALMA observations of the AB Aur disk. For scale, the VLT/SPHERE \textit{H}-band scattered light image at 1.6 $\mu$m is shown in Figure~\ref{fig:1}a (\citealp{boccaletti2020}; \citealp{speedie2024} reduction). 
Figure~\ref{fig:1}b presents our ALMA $^{12}$CO $J=2-1$ integrated intensity map, collapsed uniformly over velocities $\pm 6$ km/s about the systemic velocity \cite[$v_{\rm sys}=5.85$ km/s; ][]{tang2012-abaur-envelope, speedie2024}. 
We find that the disk extends to $r \sim 10''$ ($r \sim 1600$ au) in $^{12}$CO emission, placing it {among} the largest protoplanetary disks known both physically and on the sky {\cite[e.g.,][]{berghea2024-dracula, monsch2024-dracula}}. 
Spiral structure at all radii and azimuth is visible directly in the integrated intensity map. The ratio of the disk emission extent ($\sim20''$) to the synthesized ALMA beam ($\sim0.2''$) is $\sim100$. 

Figure~\ref{fig:1}c shows our SO $5_6-4_5$ peak intensity map, generated over a velocity range $\pm 2.5$ km/s about $v_{\rm sys}$.  
We observe a ring of SO emission between $r = 1.0'' - 2.0''$ on the sky. The ring exhibits substructure, namely 
a brightness peak in the northeast. 
Measuring the azimuthal extent of the brightness peak as the region where the peak intensity is above $10$ mJy/beam ($70\%$ of the maximum), we find a position angle range of $32^{\circ}-82^{\circ}$ east of north. The maximum SO intensity inside this region is $14.5$ mJy/beam, which is a relative increase of $150\%$ from the azimuthal average intensity of $5.7$ mJy/beam measured  throughout the rest of the ring. 
We confirmed that this brightness peak is not an artifact of continuum subtraction (see Figure \ref{appfig:SO-wcont} in Appendix \ref{app:contsub-SO}).

\section{{The Disk and Exo-Disk Components of $^{12}$CO}} \label{subsec:disentangling}

In this section and \S\ref{subsec:analytic-modeling}, we present a position-velocity (PV) analysis of the $^{12}$CO $J=2-1$ and SO $J_{N}=5_6-4_5$ emission. 
{We begin in \S\ref{subsec:radial-PV-diagrams} with radial PV diagrams to reveal the complexity in the $^{12}$CO data.} 
In \S\ref{subsec:kep-antikep-masks} we kinematically disentangle the $^{12}$CO into its {disk-like} and {non-disk-like} components, in preparation to model the latter component as infall in \S\ref{subsec:analytic-modeling}. 

\subsection{{Radial Position-Velocity Diagrams}} \label{subsec:radial-PV-diagrams}

\begin{figure*}
\includegraphics[width=\textwidth]{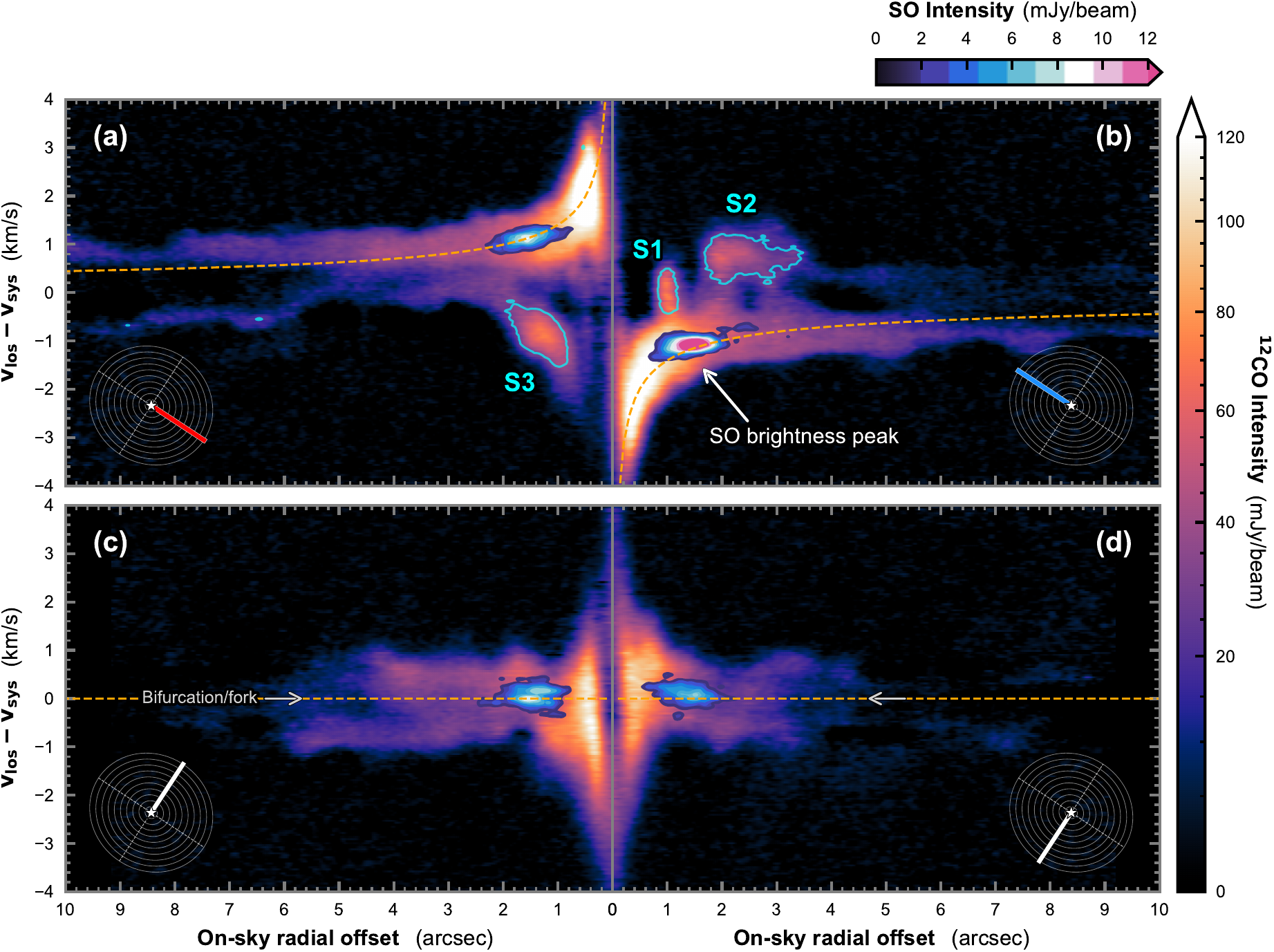}
\caption{\textbf{Position-velocity (PV) slices in ALMA $^{12}$CO $J=2-1$
emission along the disk major and minor axes.} PV slices in the ALMA SO
$J_{N}=5_6-4_5$ emission are overplotted as filled contours in
increments of $2\sigma$ starting at $3\sigma$, {where $\sigma=0.63$
mJy/beam.} The orientation of each slice is indicated in the bottom
corner of each panel: 
\textbf{(a)} Redshifted (west) major axis. 
\textbf{(b)} Blueshifted (east) major axis. 
\textbf{(c)} Northern minor axis. 
\textbf{(d)} Southern minor axis.
{Keplerian rotation is denoted by orange dashed lines, accounting} 
for the mass of the central star only. 
`Out-of-plane' spiral arms S1, S2 and S3 identified by \citet{tang2012-abaur-envelope} are labeled, and encircled by cyan contours {from the anti-Keplerian weighted $^{12}$CO cube (see next figures).} 
{In panels (c) and (d), grey arrows point to a bifurcation pattern we attribute to interferometric filtering (\S\ref{sec:data}).} 
\revv{An animation of radial position-velocity slices in ALMA $^{12}$CO $J=2-1$ and SO
$J_{N}=5_6-4_5$ emission is \href{https://figshare.com/articles/media/Mapping_the_merging_zone_of_late_infall_in_the_AB_Aur_planet-forming_system/28205066?file=51680915}{available} in the online Journal. The animation sequence pans through all azimuthal angles in $0.5\degr$ steps following the rotation direction of the disk.}
}
\label{fig:2}
\end{figure*}

Figure~\ref{fig:2} presents radial PV diagrams of $^{12}$CO $J=2-1$ and SO $J_{N}=5_6-4_5$ emission extracted along the disk's major and minor axes. 
The SO is plotted over the $^{12}$CO as filled contours. 
The PV slices are taken assuming a geometrically thin disk with inclination $i=23.2^{\circ}$ \cite[from the continuum; ][]{tang2012-abaur-envelope, tang2017-abaur12COspirals} and position angle ${\rm P.A.}=236.7^{\circ}$ \cite[where P.A. is measured east of north to the redshifted major axis; ][]{speedie2024}. We used \texttt{pvextractor} \citep{robitaille2018-glue-pvextractor} for all PV slicing throughout this work.

{Along the major axis (top row of Figure~\ref{fig:2}),} 
the projection of the disk's rotation is maximized, and we observe substantial excess $^{12}$CO emission cleanly distinct in velocity from the bulk rotating component. 
For comparison, we overlay dashed orange curves of Keplerian rotation {in a razor thin disk}, following $v_{\rm LOS} - v_{\rm sys} = v_{\rm Kep} \, \sin{i} \, \cos{\phi}$, where $v_{\rm Kep} = (G M_{\star} / r)^{1/2}$, 
the stellar mass $M_{\star} = 2.23 \, M_{\odot}$, 
the systemic velocity $v_{\rm sys} = 5.86$ km/s \citep{speedie2024}, 
and $\cos{\phi}=1$ on the redshifted major axis.

Three strong `blobs' are identifiable within the excess $^{12}$CO emission in the top row of Figure~\ref{fig:2} {(labeled S1 to S3)}.
They correspond to radial cross sections through three ``out-of-plane'' spiral structures first identified by \citet{tang2012-abaur-envelope} in $^{12}$CO $J=2-1$ observations with the Plateau de Bure interferometer (PdBI).  
Following \citet{tang2012-abaur-envelope}, we hereafter refer to them as S1, S2, and S3. They have additionally been observed in the same transition with NOEMA by \citet{riviere2020-rosetta1}. To our knowledge, we present the first ALMA detection.

Note that the Keplerian rotation curves do not account for the disk's contribution to the gravitational potential. The disk component of $^{12}$CO emission shows substantial apparent super-Keplerian rotation at large radii in the major axis PV diagrams, indicative of significant self-gravity and a high disk-to-star mass ratio. We keep to qualitative statements here because these major axis slices were taken along a straight line, which is equivalent to assuming the $^{12}$CO emission surface is the disk midplane. For now we note that the surface is almost certainly elevated, and thus the slices are likely sampling slightly away from the loci of $v_{\rm LOS}$ maxima. A quantitative kinematic analysis to fit for the disk mass should be done \cite[e.g. ][]{veronesi2021-elias227, lodato2023-gmlup-imlup, martire2024-maps, andrews2024-behemoth}, 
{though complexities in the $^{12}$CO emission, the non-axisymmetric nature of the disk and its low inclination complicate the extraction of a $^{12}$CO emission layer.} 
We tender that challenge to future work.

Along the minor axis (bottom row of Figure~\ref{fig:2}), 
we notice how the $^{12}$CO emission appears to bifurcate (or fork) into two bands above and below the horizontal line of $v_{\rm LOS}-v_{\rm sys}=0$ at large offsets, starting around $r \gtrsim 3.5''$. We understand this to be a result of interferometric spatial filtering in the central channels (\S\ref{sec:data}).

\subsection{{Keplerian \& Anti-Keplerian Masking}} \label{subsec:kep-antikep-masks}

Next, we disentangle the bulk rotating component of $^{12}$CO emission from all its {non-disk-like}  emission, in a fashion similar to \citet[][see their Appendix C]{huang2021-MAPS-GMAur}. 
We generate a Keplerian mask \citep{teague2020-keplerianmask-zenodo} to kinematically encompass the disk, and an ``anti-Keplerian'' mask with opposite Boolean values to encompass {any} emission kinematically inconsistent with the disk. 
We additionally spectrally smooth each mask, i.e., convolve the top hat spectrum in each spatial pixel with a Gaussian, to mitigate edge artifacts \cite[c.f. Figure 17 of][]{huang2021-MAPS-GMAur}, 
which also introduces some overlap in their spectra. 
Two new versions of the $^{12}$CO $J=2-1$ ALMA image cube are then generated by multiplication with each of the Keplerian and anti-Keplerian masks. In this way, the masks serve effectively as 3D weighting functions.  
Further details are provided in Appendix \ref{app:masks-kep-antikep} and Figure~\ref{appfig:masks}. 
In the remainder of this letter, we refer to the emission in the Keplerian-weighted and anti-Keplerian-weighted $^{12}$CO cubes as the ``disk'' and ``exo-disk''\footnote{As in `outside of disk', in the same sense as `exo-terrestrial' or `exo-solar planet'.} components, respectively.

\begin{figure*}
\begin{center}
    \includegraphics[width=0.9\textwidth]{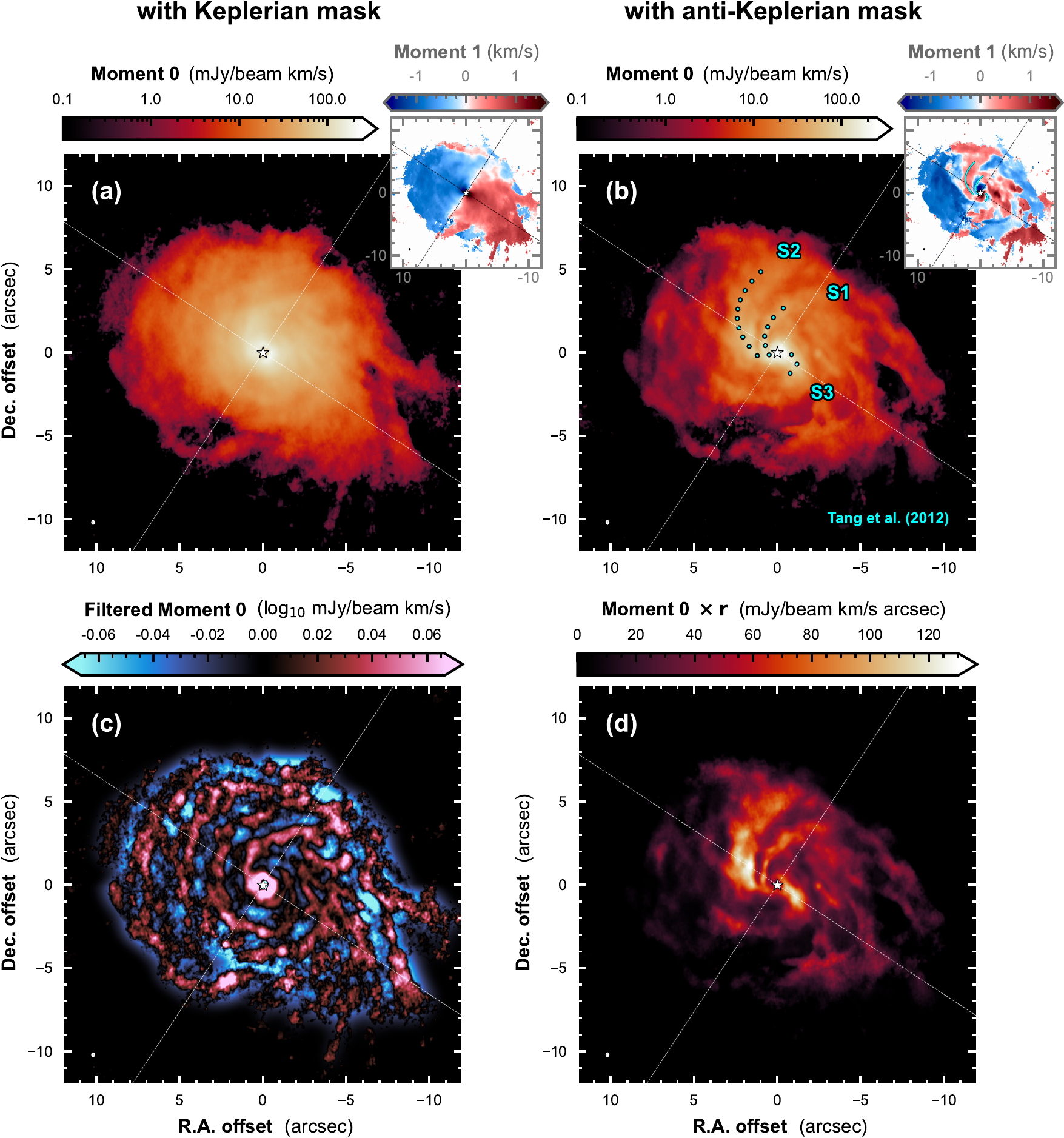}
\end{center}
\caption{\textbf{ALMA $^{12}$CO $J=2-1$ moment maps after disentangling the disk and exo-disk emission.} 
\textbf{(a)} Integrated intensity map of $^{12}$CO after weighting with a spectrally-smoothed Keplerian mask.
\textbf{(b)} The same, but after weighting with an anti-Keplerian mask generated with opposite Boolean values prior to spectral smoothing.  
Views of the masks are provided in Appendix Figure~\ref{appfig:masks}. 
The cyan points trace S1, S2 and S3 from Table 4 of \citet{tang2012-abaur-envelope}. 
Panels (a) and (b) are shown on a $\log_{10}$ colorscale to compare with Figure~\ref{fig:1}b. 
Intensity-weighted velocity maps are shown in the insets. 
\textbf{(c)} High-pass filtered integrated intensity map of the Keplerian-weighted $^{12}$CO cube in panel (a), highlighting the architecture of spiral structure within the disk. 
\textbf{(d)} The same as panel (b), but scaled by $r$ (on-sky) and presented on a linear colorscale. 
}
\label{fig:3}
\end{figure*}

Figure~\ref{fig:3} shows {moment} maps collapsed from the Keplerian and anti-Keplerian $^{12}$CO $J=2-1$ image cubes. 
We emphasize that the masking process is not perfect, but {is meaningfully effective} considering the complexity of the $^{12}$CO emission (Fig.~\ref{fig:2}). 
{Beginning with the left column, Figure~\ref{fig:3}a shows the integrated intensity map of the map of the Keplerian-weighted $^{12}$CO, or disk component. The inset panel 
shows the corresponding}  
intensity-weighted velocity map, demonstrating that the emission {isolated by the Keplerian mask} is at least broadly recognizable as disk-like, with a redshifted and blueshifted side \cite[compare to Figure 3 of ][]{riviere2020-rosetta1}. 
Figure~\ref{fig:3}c shows a {high-pass filtered version of Figure~\ref{fig:3}a}, {made} with a radially expanding Gaussian kernel\footnote{\url{github.com/jjspeedie/expanding_kernel}}  
{using the same parameter setting as \citet{speedie2024}}. 
A rich architecture of spiral structure exists in the disk component, {consistent with} ongoing gravitational instability \cite[e.g., ][]{dipierro2014, hall2019-temporalGIspiralsALMA}.

In the right column of Figure~\ref{fig:3}, we show {moment maps of} the anti-Keplerian-weighted $^{12}$CO cube, {or exo-disk component}. 
{Figure~\ref{fig:3}b presents the integrated intensity map, where} points along the spiral structures S1, S2, and S3 placed by \citet[][their Table 4]{tang2012-abaur-envelope} are overplotted in cyan for reference. 
Figure~\ref{fig:3}d shows the integrated intensity map $r$-scaled with a linear colorbar. The exo-disk component is comprised predominantly of coherent spirals S1, S2 and S3, with additional `whisps' of emission (particularly in the west and south).  
The corresponding intensity-weighted velocity map (Figure~\ref{fig:3}b inset) is 
{markedly non-disk-like,} with 
regions of emission at `opposite' line-of-sight velocity to the disk.

We note that the imprint of S1 is visible in the \textit{disk} component (compare Figure~\ref{fig:3}c and d). 
{In other words, some of its emission remains within the Keplerian mask.} As we'll discuss in the next section, this is because emission from S1 is not sufficiently offset in line-of-sight velocity from the {disk} component \textit{along its entire} trajectory (not just at the major axis as we glimpsed in Figure~\ref{fig:2}b). On the other hand, S2 and S3 are kinematically distinct from the disk at almost all their spanned azimuths.

\section{Modeling the Exo-disk Component as Infall } \label{subsec:analytic-modeling}

We focus on the coherent exo-disk spiral structures S1, S2, and S3, and show that their trajectories (in both RA-Dec and PV space) can be reproduced with the analytic accretion flow model of \citet{mendoza2009}, 
using the numerical implementation by \citet{pineda2020-natast}. 
Interpreting the exo-disk emission as infall is well motivated: {observations of AB Aur's kilo-au environment are highly suggestive of interaction with surrounding material \cite[e.g.,][]{grady1999-hst}}. {See more details in Figure \ref{appfig:environment} of Appendix \ref{app:environment}}. 
The next subsection outlines the scheme of the model, with our results presented in 
\S\ref{subsubsec:analytic-solutions}.

\subsection{The Pineda Implementation of the Mendoza Streamline Model (PIMS)} \label{subsubsec:pims}

The \citet{mendoza2009} model considers a steady-state ballistic accretion flow from a rigidly rotating sphere 
of finite radius $r_{\rm 0}$ (representing the inner edge of a giant molecular cloud) 
toward a central object with mass $M$, located at the origin of coordinates.  
The angular frequency of the sphere's rigid-body rotation is $\Omega$. 
In spherical coordinates, a fluid `particle' is initiated on the sphere at an azimuthal angle $\phi_{\rm 0}$ and a polar angle $\theta_{\rm 0}$ (where $\theta = 90^{\circ}$ defines the {equatorial plane}). It can also be given an initial radial velocity, $v_{r, 0}$. 
The particle has specific angular momentum $h$ {about the origin} 
inherited from its initial location, where $h$ is distributed on the sphere according to {$h = h_{\rm 0} \, \sin{\theta}$ assuming $h_{\rm 0} = r^{2}_{\rm 0} \, \Omega$}. 
The particle's specific energy $E$ is the sum of its 
specific kinetic energy, centrifugal potential energy, and gravitational potential energy: {\begin{equation}
    E = \frac{1}{2}v^2_r + \frac{1}{2}\frac{h^2}{r^2} - \frac{GM}{r} = \frac{1}{2}v^2_{r,0} + \frac{1}{2}\frac{h_{\rm 0}^2 \sin^2{\theta_{\rm 0}}}{r_{\rm 0}^2} - \frac{GM}{r_{\rm 0}} \, .
    \label{eqn:mendoza-eqn2}
\end{equation}} 
The quantities $h$ and $E$ are {contants of motion along each particular trajectory}. 

The trajectory of each particle (hereafter a `streamline') is a solution to Kepler's problem \citep{newton1687}, and therefore a segment of a conic section with the origin at one of the foci. 
{This means each streamline is contained in a plane, and we will later refer to this as the plane `of' a streamline.} 
The whole system is cylindrically symmetric, such that each streamline `collides' with its reflection counterpart at the {equatorial} plane. An important fundamental property of the model is that a streamline cannot spiral infinitely -- it \textit{must} hit the {equatorial} plane after traveling an azimuthal distance of $<90^{\circ}$ from its initial azimuthal location $\phi_{\rm 0}$. It comes closest to subtending this maximum azimuthal distance when initialized close to the {equatorial} plane \cite[$\cos{(\phi - \phi_{\rm 0})} = \tan{\theta_{\rm 0}}/\tan{\theta}$; Equation 8 of][]{mendoza2009}.

With the \citet{pineda2020-natast} Implementation of the \citet{mendoza2009} Streamline Model (PIMS\footnote{\url{github.com/jjspeedie/pims}}), we have 6-dimensional information --3 spatial coordinates and 3 velocity coordinates-- at every point along the streamline. 
Two rotational transformations are then applied to the coordinates to project the system onto the sky: a first to give the streamline's {equatorial} plane some inclination, and a second to give {the ascending node} some position angle\footnote{We note that the line-of-sight velocity, $v_{\rm LOS}$, is invariant under this transformation ($v_{\rm LOS}$ is independent of the position angle).}. This yields 3D spatial coordinates $\left\langle {\rm RA}, {\rm Dec}, {\rm LOS} \right\rangle$ and 3D velocity components $\left\langle v_{\rm RA}, v_{\rm Dec}, v_{\rm LOS} \right\rangle$ at every point on the streamline. 
We further compute the projected radial coordinate, $r_{\rm proj}$, as the hypotenuse of the ${\rm RA}$ and ${\rm Dec}$ coordinates. When we compare streamlines to the AB Aur data, we do so in the ${\rm RA}$-${\rm Dec}$ plane and the $r_{\rm proj}$-$v_{\rm LOS}$ (PV) plane.

We obtained 
streamline solutions for each of S1, S2 and S3 by first exploring their trajectories in radial and azimuthal PV cross sections  to build an understanding of their $v_{\rm LOS}$ behaviour as a function of $\phi$ and $r$. 
A selection of such cross sections is shown in Figure~\ref{appfig:slices} of Appendix \ref{app:slices-frames}, \rev{with animated versions available in the online article}. 
We then identified, by manual experimentation, the region of model parameter space where streamlines follow a broadly similar track in the PV plane. This procedure leverages the fact that the infall model trajectories in the $r_{\rm proj}$-$v_{\rm LOS}$ plane only take one of a few basic shapes \cite[e.g. see][for examples]{thieme2022, mori2024}. From there, we iteratively fine-tuned the parameters until the match between the streamlines and the observed trajectories 
in the RA-Dec and PV planes were visually satisfactory. 
The only parameter we did not let vary in these experiments is $M$, which we set to $M=2.90\, M_{\odot} = M_{\star} \times 130\%$ to represent both the central star and the disk \citep{speedie2024}.

In the following subsection, we present one streamline solution for each of S1, S2 and S3. The PIMS 
parameters describing streamlines are provided in {Table \ref{tab:pims_params} of} Appendix \ref{app:streamline-params}. 
We note that for each of S1 and S2, our procedure easily yielded a family of streamlines that are satisfactory matches. We confirmed each family describes a structure 
physically coherent in 3D spatial coordinates, $\left\langle {\rm RA}, {\rm Dec}, {\rm LOS} \right\rangle$ (see Figure~\ref{appfig:pyvista} in Appendix \ref{app:streamline-params}), and selected one solution to show in the main text. 
Satisfactory solutions for S3, on the other hand, were not so easily found; we nonetheless present our single best solution for S3 for discussion and for reference with \citet{tang2012-abaur-envelope}.

\subsection{Streamline Solutions for S1, S2 and S3 }  \label{subsubsec:analytic-solutions} 

\begin{figure*}
\includegraphics[width=\textwidth]{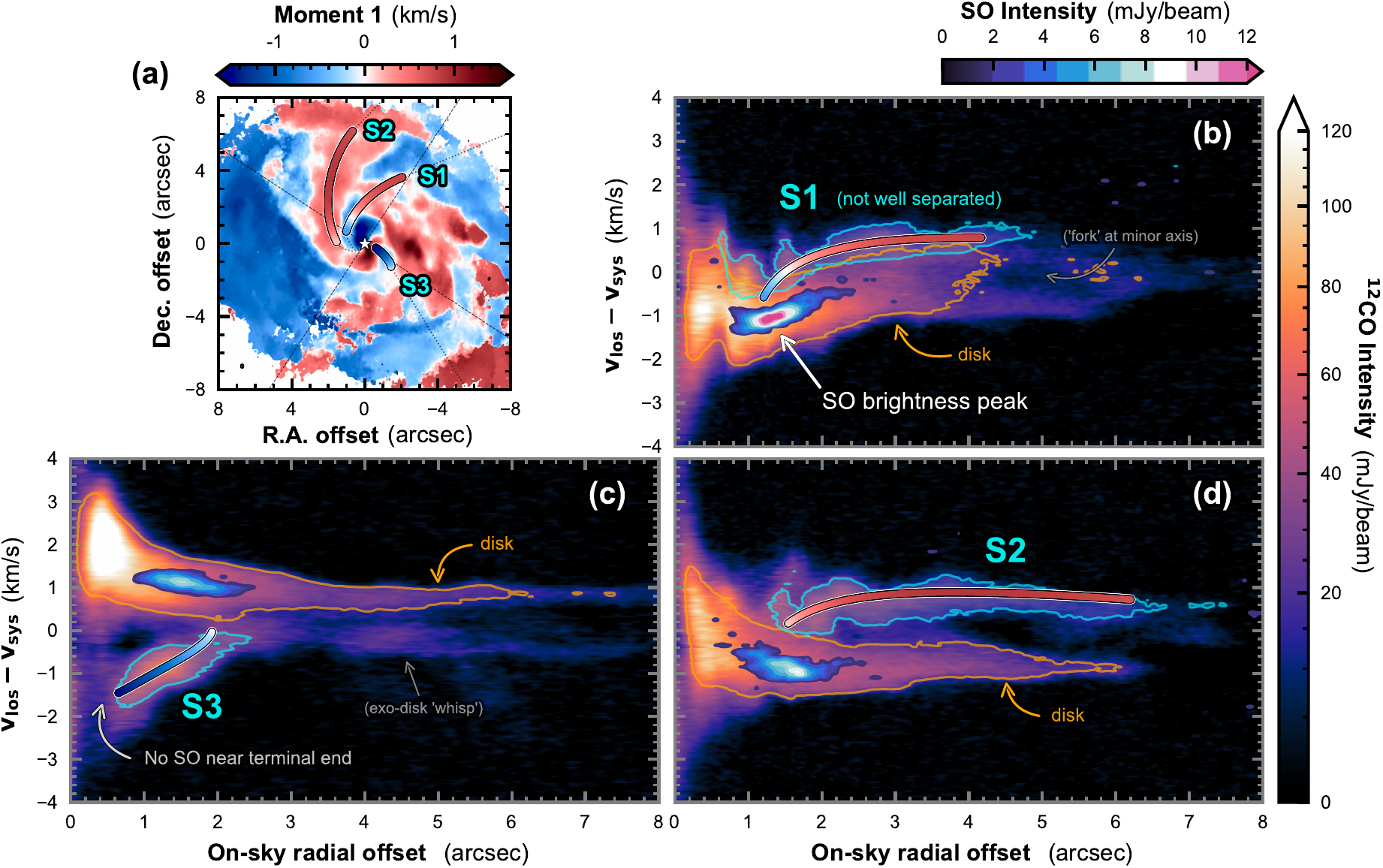}
\caption{\textbf{Modeling S1, S2 and S3 as infall.} \textbf{(a)} Intensity-weighted velocity map of the anti-Keplerian weighted $^{12}$CO $J=2-1$ image cube (as in Figure~\ref{fig:3}b inset). We overlay RA-Dec trajectories of three streamlines matching S1, S2 and S3, computed with the analytic accretion flow model of \citet{mendoza2009} implemented in \citep{pineda2020-natast}. {The colorbar for velocity applies to both the background velocity map and the model streamlines.} 
\textbf{(b,c,d)} PV slices taken \textit{along the RA-Dec trajectory} of each streamline for completely consistent comparison between model and data {(one panel per streamline)}. 
{Within a panel, multiple cubes have been sliced:} the full unweighted $^{12}$CO cube (colormap), the Keplerian weighted $^{12}$CO cube (orange contours), and the anti-Keplerian weighted $^{12}$CO cube (cyan contours). 
PV slices in the SO $J_{N}=5_6-4_5$ image cube are overplotted as filled contours in increments of $2\sigma$ starting at $3\sigma$. 
We concatenate arbitrary extensions to the slice paths so that the PV diagrams 
span $r=0''-8''$ in on-sky radial offset, shown as dotted gray lines extending from both ends of the streamline trajectories.  
Radial and azimuthal PV cross sections providing additional visualizations of S1 and S2 merging with the disk are available in Appendix Figure~\ref{appfig:slices}. 
}
\label{fig:4}
\end{figure*}

Figure~\ref{fig:4} presents our infall modeling results for each of S1, S2 and S3. 
In Figure~\ref{fig:4}a, we overlay the three PIMS streamline RA-Dec trajectories onto the $^{12}$CO anti-Keplerian moment 1 map (from Figure~\ref{fig:3}b inset). 
The streamlines fall from large radii to small radii, and are coloured by their line-of-sight velocity, $v_{\rm LOS}$. 
The S1 and S2 streamlines are predominantly redshifted (where the AB Aur disk is blueshifted) and lie in front of the disk, falling toward the disk away from us. 
The plane of S1 is approximately $20^{\circ}$ inclined with respect to the AB Aur disk midplane, and the plane of S2 is inclined by approximately $35^{\circ}$. 
S3 is blueshifted (where the disk is redshifted) and lies behind the disk, falling toward us, on a plane inclined by approximately $45^{\circ}$ to the AB Aur disk midplane. 
See Figure~\ref{appfig:pyvista} for a 3D rendering \rev{and animation} of the system geometry.

In Figure~\ref{fig:4}bcd, we overlay each of the streamline $r_{\rm proj}$-$v_{\rm LOS}$ trajectories onto PV diagrams from the $^{12}$CO image cube, where the PV slices were taken in the data cube along each of the streamlines' RA-Dec trajectories for completely consistent comparison {\cite[see Figure 9 of ][for another example of this]{huang2021-MAPS-GMAur}.} 
To help visually discern the disk and exo-disk components of the $^{12}$CO emission in this space, 
the same slices were taken through each of the Keplerian- and anti-Keplerian weighted $^{12}$CO image cubes, and overplotted with orange and cyan contours, respectively.   
Finally, we take the same slices through the SO $J_{N}=5_6-4_5$ cube, and overplot the SO emission as filled contours as well (like Figure~\ref{fig:2}). 

We find that both S1 and S2 --the observed structures-- follow a `flatline-then-dip' trajectory in the $r_{\rm proj}$-$v_{\rm LOS}$ plane as they accelerate toward the star. 
We identify the `merging zone' of S1 and S2 as the region where their emission in $^{12}$CO 
becomes indistinguishable from 
the emission from the disk, which is also where the terminal ends of our model streamlines intersect with the disk midplane in 3D spatial coordinates (Figure~\ref{appfig:pyvista}). 
Figure~\ref{fig:4}b and d 
show this from the perspective of PV slices along the streamlines themselves, and Appendix Figure~\ref{appfig:slices} shows this from the perspective of PV slices in radius and azimuth.  The merging occurs at radii $r \sim 1\arcsec -2\arcsec$ from the star and 
between position angles {$15^{\circ}-100^{\circ}$ east of north} on the sky. SO $J_{N}=5_6-4_5$ emission is observed at these same radii, peaking {within this position angle range (\S\ref{subsec:fiducial-maps})}, and at line-of-sight velocities consistent with Keplerian, indicating localized heating within the disk (discussed further in \S\ref{sec:discussion:SO}).

We note that, as mentioned in \S\ref{subsec:kep-antikep-masks}, emission from S1 is not well separated in line-of-sight velocity from the disk component along its entire trajectory (see Figure~\ref{fig:4}b), while 
emission from S2 is kinematically distinct from the disk at all its spanned azimuths {leading up to the merging zone} (Figure~\ref{fig:4}d). 
{This can be explained as a projection effect:  
Both are observed at redshifted 
line-of-sight velocities within roughly $0-1$ km/s above $v_{\rm sys}$,  
but S1 is spatially nearer to the minor axis (particularly at large radii), where the disk emission already occupies these line-of-sight velocities. Away from the minor axis (looking east), rotation projects the disk blueward of $v_{\rm sys}$, such that emission from S2 is separated in line-of-sight velocity over the course of its approach}.

{Turning to Figure~\ref{fig:4}c, S3 is observed in $^{12}$CO at blueshifted line-of-sight velocities within roughly $0-2$ km/s below $v_{\rm sys}$. Lying spatially near the western major axis, it is well separated from the disk's redshifted emission.} 
We found that the analytic infall model cannot produce a streamline that simultaneously (i) is blueshifted, (ii) is located on the redshifted side of the disk, (iii) is oriented in a counter-clockwise sense on the sky, and (iv) takes a concave-down trajectory in the $r_{\rm proj}$-$v_{\rm LOS}$ plane. 
We present a streamline that violates solely attribute (iv). 
The terminal end of the model streamline lies on the redshifted major axis at $r\sim0.5''$, both spatially and kinematically offset from the observed SO emission.

\begin{figure*}
\includegraphics[width=\textwidth]{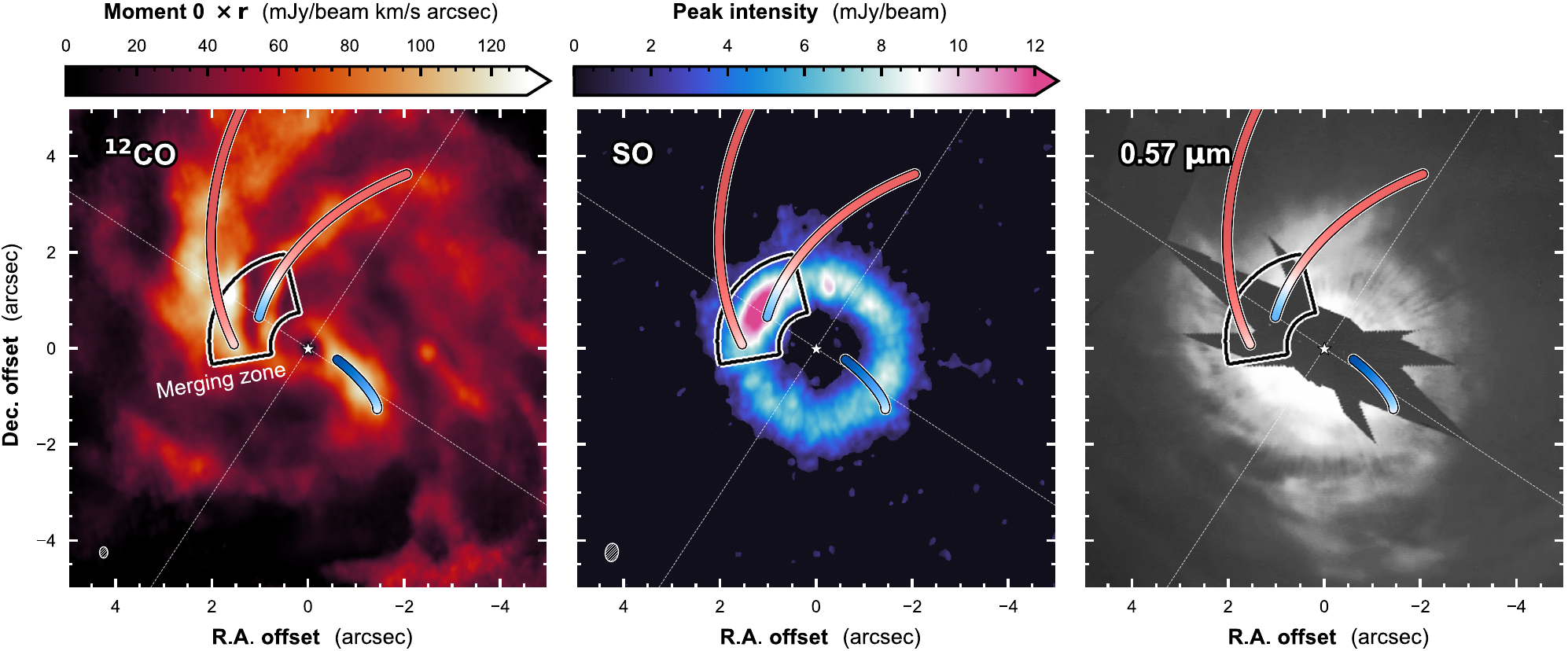}
\caption{\textbf{The merging zone of late infall in the AB Aur disk.} 
The streamline trajectories for S1, S2 and S3 from Figure~\ref{fig:4} are overlaid onto the {exo-disk} $^{12}$CO $J=2-1$ integrated intensity map (as in Figure~\ref{fig:3}d), the SO $J_{N}=5_6-4_5$ peak intensity map (as in Figure~\ref{fig:1}c) and 
HST/STIS scattered light image at $\lambda=0.57\,\mu$m \cite[Figure 3 of][reproduced with permission]{grady1999-hst}.
The terminal ends of S1 and S2 coincide with the SO $J_{N}=5_6-4_5$ brightness peak. 
A rendering of the disk+streamline geometry in 3-dimensional space is available in Appendix Figure~\ref{appfig:pyvista}. 
}
\label{fig:5}
\end{figure*}

\section{Discussion} \label{sec:discussion}

Figure~\ref{fig:5} provides a summary of our main result. 
We have applied a simple ballistic accretion flow model \cite[used by previous works to analyze streamers on cloud or envelope scales, e.g.,][]{pineda2020-natast, valdiviamena2022-peremb50, garufi2022-hltau-dgtau, valdiviamena2023-b5, flores2023-edisk-irs63} onto \textit{disk-scales}, in a system with multiple infalling disk-scale streamers and several embedded planet candidates. 
Figure~\ref{fig:5}c shows that S1 and S2 are seen in $0.57\, \mu$m HST/STIS imaging \citep{grady1999-hst}, consistent with being in front of the disk. 
We identify the `merging zone' of S1 and S2 as the region where their emission in $^{12}$CO becomes indistinguishable from the emission from the disk (Figure~\ref{fig:4} \& \ref{appfig:slices}), which is also where our model streamlines intersect with the disk midplane in 3D spatial coordinates (Figure~\ref{appfig:pyvista}). In this same region, we observe a brightness peak in SO $J_{N}=5_6-4_5$ emission. We highlight that the merging zone is in close proximity to candidate sites of planet formation.

\subsection{What are the origins of late infall in the AB Aur system?} \label{sec:discussion:origins}

Late-stage infall events may be the result of low-velocity cloud fragments being gravitationally captured by the star, and accreted through a Bondi-Hoyle Lyttleton (BHL) type process \citep{bondi1952, edgar2004-review}.  Zooming out to AB Aur's kilo-au environment, optical images with the University of Hawaii 2.2 m Telescope show an arc-shaped structure or ``reflection nebula'' surrounding AB Aur from the east to the south \citep{grady1999-hst, nakajima-golimowski1995-palomar}. 
This arc structure is hypothesized to be part of a cloudlet 
unrelated to the early protostellar collapse phase -- specifically, a part that was \textit{not} captured by the gravity of AB Aur, and instead was bent during a flyby encounter onto an arc-shaped (hyperbolic) trajectory \citep{dullemond2019-cloudlet, kuffmeier2020-late-encounters, gupta2023-reflection-nebulae}. Such a scenario, but playing out around DG Tau, was recently demonstrated with hydrodynamic simulations by \citet{hanawa2024-cloudlet-capture-DGtau}. 

In Figure \ref{appfig:environment} (Appendix \ref{app:environment}) we report new but very tentative evidence supporting this hypothesis in AB Aur. The closest projected separation of the reflection nebula, $\sim1500$ au or $\sim10''$, falls within the field of view of our ALMA observations ($38'' \times 38''$). In $^{12}$CO emission, we observe a faint `tail-like' emission structure extending from the disk to the south, bridging the disk to a source outside the primary beam FOV.  
The tail appears co-spatial with the reflection nebula on the sky, hinting to a physical connection and suggesting that some part of the cloudlet has been gravitationally captured. 
Further notes on the tail structure are provided in Appendix \ref{app:environment}. 
We stress this detection is tentative as the relative response of the ALMA 12m antenna sensitivity tapers to the $20\%$ level at the edge of our imaged FOV (\S\ref{sec:data}). Follow-up kinematic observations probing a larger region of sky and with sensitivity to those larger spatial scales are needed. The connection, if any, between the `tail-like' emission at the southern edge of the FOV and the streamers observed \textit{within} the confines of the disk --S1, S2 and S3-- is unclear.

\subsection{What are the consequences of late infall in the AB Aur system?} \label{sec:discussion:consequences}

The consequences of infall are fundamental and wide-reaching \cite[e.g.,][]{kuffmeier2023-rejuvenating-infall, pelkonen2024-BHL, winter2024-BHL}. 
In this section we focus our lens on angular momentum transport and disk substructure. 

Material infalling into a pre-existing disk must trigger some form of angular momentum redistribution for the new mass to be `assimilated' into the disk
\cite[e.g.,][]{kratter-lodato2016}. 
Numerical simulations have shown gravitational instabilities to be triggered and/or spiral structure to be formed as a consequence of infall \cite[e.g., ][]{harsono2011-GI-infall, lesur2015-spiral-accretion, hennebelle2017-infall-spirals, kuffmeier2018-infall-instabilities}. 
{Indeed, infall has been previously proposed to trigger GI in Elias 2-27 \citep{paneque2021-elias27}}. 
In an evolved system like AB Aur, an influx of mass may have been necessary to
acquire and maintain a high disk-to-star mass ratio \citep{fukagawa2004, hall2019-temporalGIspiralsALMA}. 
If gravitationally unstable disks only exhibit prominent spiral structure for a limited time after the onset of GI, 
as suggested by recent numerical simulations 
\citep{rowther2024-shortlivedGI}, 
then we may infer that gravitational instability must have been triggered --e.g. by infall-- relatively recently.

In addition to gravitational instabilities, simulations have shown infall to launch the Rossby wave instability (RWI), resulting in the ready formation of vortices and pressure bumps where material is deposited \citep{bae2015-infall-rwi, kuznetsova2022-infall-pressurebumps-vortices}. 
This expectation appears to be borne out in the AB Aur system, which displays a ring and azimuthal asymmetry in mm continuum emission at $r\sim1''$ \cite[e.g., Figure~\ref{appfig:SO-wcont},][]{tang2017-abaur12COspirals}, just interior to the merging zone identified here.

The sheer size of the AB Aur disk ($r\sim1600$ au in $^{12}$CO; \S\ref{subsec:fiducial-maps}) may also be ascribed to late infall, as extraordinarily large disks are statistically more likely to have experienced infall with high angular momentum \citep{kuffmeier2023-rejuvenating-infall}. 
Misalignment between the angular momentum vector of infalling material and that of the inheriting disk 
can induce warping  
\citep{sai2020-L1489-warp, kuffmeier2021-infall-misalignments, dullemond2022-kimmig-zanazzi, kuffmeier2024-reorientation}, with a severity that depends on the mass ratio of the two. 
Some degree of warping appears to be present in the AB Aur disk, 
based on the non-axisymmetry of the background velocity field traced by $^{13}$CO and C$^{18}$O \citep{speedie2024}.

\subsection{How do we interpret the SO emission morphology? } \label{sec:discussion:SO}

SO emission in the AB Aur system has been observed in five transitions with three facilities to date:  
SO $3_4-2_3$ and $5_6-4_5$ with the IRAM 30m telescope \cite[][not spatially resolved]{fuente-2010-iram-abaur}; 
SO $5_6-4_5$ with NOEMA in CD configuration \cite[][spatially resolved for the first time]{pacheco-vazquez2016-SO}; 
SO $5_5-4_4$ and $5_6-4_5$ with NOEMA in AC configuration \citep{riviere2020-rosetta1}; and 
SO $5_6-4_5$, $6_5-5_4$ and $6_7-5_6$ with ALMA \citep{dutrey2024-abaur-SO}. 
In the latter three works, where SO $5_6–4_5$ is spatially resolved, a brightness peak within a ring of emission is found in the same location reported here. 
In this section, we discuss its production and survival.

The merging of S1 and S2 will be accompanied by shock waves if the infalling flow is supersonic. This is typically the case in infall, with the speed of sound being $\lesssim0.2$ km/s at $r\gtrsim100$ au. While we only measure line-of-sight velocities of S1 and S2 in the data (Figure~\ref{fig:4}), we can estimate their incident velocity using the known velocity components of the model streamlines, which are listed in Appendix Table \ref{tab:pims_params}. Their three velocity components (inward, clockwise, and toward the midplane) range $0.7-4.1$ km/s, consistent with `low-velocity' (usually quoted to mean $2$ km/s) shocks.

Within shocks, the kinetic energy of the infalling material is converted into thermal energy. 
Dust grains are heated by conduction from the shocked gas, and by increased aerodynamic drag if the gas is decelerated by pressure forces \cite[e.g.,][]{miura2017-desorption-SO}. If the resulting dust temperature is sufficiently high (of order $40-70$ K), SO can be produced directly via thermal sublimation of SO ice mantles \cite[e.g.,][]{{vangelder2021-shocks-sulphur}}. 
Within the SO ring in the AB Aur system, \citet{riviere2020-rosetta1} and \citet{pacheco-vazquez2016-SO} estimate an average gas kinematic temperature of $\sim 39$ K from H$_2$CO and SO multi-transition analysis. 
{If aerodynamic heating operates alongside conductive heating, the gas kinematic temperature may represent a lower bound for the dust temperature.}

Thermal sublimation is most efficient for small grains ($<0.1 \, \mu$m) because they are more significantly heated, due to their lower emissivities \citep{miura2017-desorption-SO}. 
    They also likely represent the largest contribution to the dust-grain surface area \cite[e.g.][]{mathis1977}. 
HST/STIS imaging of AB Aur at $0.57 \, \mu$m wavelengths \cite[tracing grain sizes $\lambda/2\pi \sim 0.09\mu$m, e.g.,][]{kruegel2003} confirms such small grains are present high in the disk atmosphere, and with a surface brightness distribution following S1 and S2 \cite[Figure~\ref{fig:5}c;][]{grady1999-hst}.

Gas-phase formation of SO after the sputtering of s-CH$_4$ and s-H$_2$S may be an alternative possibility, in which case we should observe H$_2$CO and SH emission coincident with SO \citep{vangelder2021-shocks-sulphur}. Indeed, the former has been seen by \citet{pacheco-vazquez2016-SO} and \citet{riviere2020-rosetta1}. On the other hand, sputtering has been found to only become relevant at larger shock velocities \cite[$>10$ km/s;][]{aota2015}.

Regardless of which production mechanism, if SO \textit{is} produced at the merging zone of S1 and S2, it would need to survive (only) {$\sim 3500$ years} to be circularized, based on the orbital {period timescale, $P= r^{3/2} \, M_{\star}^{-1/2}$,} at the outermost radius of the SO ring ($r\sim 300$ au). 
Chemical modeling of the AB Aur system by \citet{pacheco-vazquez2016-SO} suggests SO molecules are depleted in less than $0.1$ Myr (at densities $>10^{7}$ cm$^{-3}$ and assuming a fixed dust temperature of $45$ K), which is a {$\sim30\times$} longer timescale. 
Looking toward other systems, there are cases where SO emission associated with the merging zone of infall has been observed as only a clump, rather than a ring, despite shorter circularization timescales: for example, DR~Tau \cite[][$350$ years]{huang2024-drtau-so}, DG~Tau and HL~Tau \cite[][$600-1200$ years]{garufi2022-hltau-dgtau}. We may conclude there are different physical conditions in these systems, or deeper observations may help to further illuminate the SO emission morphologies.

\section{Conclusions} \label{sec:conclusions}

We have presented deep and high spectral resolution ALMA observations (ID: 2021.1.00690.S) of $^{12}$CO $J=2-1$ and SO $J_{N}=5_6-4_5$ emission toward the Class II YSO AB Aur. 
In $^{12}$CO, we mapped three infalling `streamers' spatially coincident with the disk, constrained their physical 3D geometries, and showed they can be modeled down to where they merge with the disk.  
A summary of our approach and findings is as follows.

\begin{itemize} \item Using a combination of Keplerian and anti-Keplerian masking, we isolate the {disk-like} $^{12}$CO emission from {non-disk-like emission, which we term} the `exo-disk' component (\S\ref{subsec:kep-antikep-masks}). 
    \item The disk emission of $^{12}$CO extends up to $\sim 1600$ au in radius, and exhibits global spiral structure consistent with gravitational instability (Figure~\ref{fig:3}). 
\item Within the exo-disk emission, we re-detect the three $^{12}$CO spiral structures  S1, S2 and S3, first observed by \citet{tang2012-abaur-envelope} with the PdBI. 
    Using the ballistic accretion flow  model of \citet{mendoza2009} and \citet{pineda2020-natast}, we reproduce the on-sky and position-velocity trajectories of S1, S2 and S3 in detail (Figure~\ref{fig:4}). \item Our modeling results indicate that S1 and S2 are infalling onto the disk from in front (Appendix Figure~\ref{appfig:pyvista}), consistent with their traces in HST/STIS imaging of $0.57\,\mu$m scattered light (Figure \ref{fig:5}). \item We identify the `merging zone' of S1 and S2 as the region where their emission {becomes indistinguishable with that} of the disk in PV space (Figure~\ref{fig:4} \& ~\ref{appfig:slices}), and where their model streamlines intersect with the disk midplane (Figure~\ref{appfig:pyvista}).
    \item The merging zone of S1 and S2 lies {$15^{\circ}-100^{\circ}$ east of north} and $150-300$ au in radius on the sky, coincident with where a ring of SO emission exhibits a brightness 
{asymmetry peaking at $2.5\times$} 
    its azimuthal average level (Figure~\ref{fig:5}). 
\end{itemize}

We consider infall to be a strong explanation for the observed gravitational instability and implied high disk mass in the AB Aur system \cite[\S\ref{sec:discussion:consequences};][]{speedie2024}. The origin of the infalling material in this system remains an open question (\S\ref{sec:discussion:origins}), 
though kinematic observations probing larger spatial scales may affirm the cloudlet capture hypothesis \cite[Appendix Figure~\ref{appfig:environment};][]{dullemond2019-cloudlet}. 
Detailed kinematic investigations of how late-stage infall {interfaces with the disk} should be undertaken in a larger number of planet-forming systems {to understand its influence on} disk structure and the local planet formation environment.

\section{Acknowledgments} 
{We thank our referee for providing a helpful review of the manuscript, including thoughtful suggestions and questions.}
We are grateful to Michael Kuffmeier, Álvaro Ribas, Adele Plunkett, {Andrew Winter, Myriam Benisty,} Pablo Rivière-Marichalar, Asunción Fuente, Takayuki Muto, Yifan Zhou and Sahl Rowther for enlightening discussions, {and to Giulia Perotti for valuable comments on the manuscript.} 
J.S. thanks Ryan Loomis, Sarah Wood and Tristan Ashton at the North American ALMA Science Center (NAASC) for providing science support and technical guidance on the ALMA data as part of a Data Reduction Visit to the NAASC, which was funded by the NAASC. The reduction and imaging of the ALMA data was performed on NAASC computing facilities. 

J.S. acknowledges financial support from the Natural Sciences and Engineering Research Council of Canada (NSERC) through the Canada Graduate Scholarships Doctoral (CGS D) program. 
R.D. acknowledges financial support provided by the Natural Sciences and Engineering Research Council of Canada through a Discovery Grant, as well as the Alfred P. Sloan Foundation through a Sloan Research Fellowship. 
C.L. and G.L. acknowledge funding from the European Union’s Horizon 2020 research and innovation programme under the Marie Sklodowska-Curie grant agreement \# 823823 (RISE DUSTBUSTERS project). C.L. acknowledges funding from UK Science and Technology research Council (STFC) via the consolidated grant ST/W000997/1.
B.V. acknowledges funding from the ERC CoG project PODCAST No 864965. 
Y.W.T. acknowledges support through NSTC grant 111-2112-M-001-064- and 112-2112-M-001-066-.
J.H. was supported by JSPS KAKENHI Grant Numbers 21H00059, 22H01274, 23K03463. 
J.C. acknowledges support from the National Natural Science Foundation of China under grant No. 12233004 and 12250610189. 
D.S.-C. is supported by an NSF Astronomy and Astrophysics Postdoctoral Fellowship under award AST-2102405.

This paper makes use of the following ALMA data: ADS/JAO.ALMA\#2021.1.00690.S. ALMA is a partnership of ESO (representing its member states), NSF (USA) and NINS (Japan), together with NRC (Canada), MOST and ASIAA (Taiwan), and KASI (Republic of Korea), in cooperation with the Republic of Chile. The Joint ALMA Observatory is operated by ESO, AUI/NRAO and NAOJ. The National Radio Astronomy Observatory is a facility of the National Science Foundation operated under cooperative agreement by Associated Universities, Inc.

Based on data products created from observations collected at the European Organisation for Astronomical Research in the Southern Hemisphere under ESO programme 0104.C-0157(B). 
This work has made use of the SPHERE Data Centre, jointly operated by OSUG/IPAG (Grenoble), PYTHEAS/LAM/CESAM (Marseille), OCA/Lagrange (Nice), Observatoire de Paris/LESIA (Paris), and Observatoire de Lyon. 

This research used the Canadian Advanced Network For Astronomy Research (CANFAR) operated in partnership by the Canadian Astronomy Data Centre and The Digital Research Alliance of Canada with support from the National Research Council of Canada the Canadian Space Agency, CANARIE and the Canadian Foundation for Innovation. 

Based on observations made with the NASA/ESA Hubble Space Telescope, obtained from the Data Archive at the Space Telescope Science Institute, which is operated by the Association of Universities for Research in Astronomy, Inc., under NASA contract NAS5-26555. These observations are associated with program \#8065. 

Herschel is an ESA space observatory with science instruments provided by European-led Principal Investigator consortia and with important participation from NASA. 
PACS has been developed by a consortium of institutes led by MPE (Germany) and including UVIE (Austria); KU Leuven, CSL, IMEC (Belgium); CEA, LAM (France); MPIA (Germany); INAF-IFSI/OAA/OAP/OAT, LENS, SISSA (Italy); IAC (Spain). This development has been supported by the funding agencies BMVIT (Austria), ESA-PRODEX (Belgium), CEA/CNES (France), DLR (Germany), ASI/INAF (Italy), and CICYT/MCYT (Spain). 
SPIRE has been developed by a consortium of institutes led by Cardiff University (UK) and including Univ. Lethbridge (Canada); NAOC (China); CEA, LAM (France); IFSI, Univ. Padua (Italy); IAC (Spain); Stockholm Observatory (Sweden); Imperial College London, RAL, UCL-MSSL, UKATC, Univ. Sussex (UK); and Caltech, JPL, NHSC, Univ. Colorado (USA). This development has been supported by national funding agencies: CSA (Canada); NAOC (China); CEA, CNES, CNRS (France); ASI (Italy); MCINN (Spain); SNSB (Sweden); STFC, UKSA (UK); and NASA (USA).

\vspace{5mm}
\facilities{ALMA, VLT, Herschel, HST}

\software{\texttt{astropy} \citep{astropy:2013, astropy:2018}, 
          \texttt{bettermoments} \citep{teague2018-bettermoments, teague2018-robust-linecentroids}, 
          \texttt{CASA} \citep{mcmullin2007-casa}, 
          \texttt{CMasher} \citep{2020-cmasher}, 
          \texttt{gofish} \citep{teague2019-gofish-joss},  
          \texttt{keplerian\_mask} \citep{teague2020-keplerianmask-zenodo}, 
          \texttt{matplotlib} \citep{Hunter:2007},
          \texttt{numpy} \citep{harris2020array}, 
          \texttt{pandas} \citep{2011-mckinney-pandas},
          \texttt{pvextractor} \citep{robitaille2018-glue-pvextractor}, 
          \texttt{pyvista} \citep{sullivan2019pyvista},            
          \texttt{scipy} \citep{2020SciPy-NMeth}. 
          }

\section*{Data Availability}
All ALMA data products presented in this work are available through the 
\href{https://www.canfar.net/en/docs/digital_object_identifiers/}{CANFAR Data Publication Service} 
at \dataset[doi:10.11570/24.0098]{https://doi.org/10.11570/24.0098}. 
This includes final reduced and calibrated ALMA measurement sets, image cubes, masks and moment maps. 
The raw ALMA data are publicly available via the ALMA archive 
\url{https://almascience.nrao.edu/aq/} under project ID  2021.1.00690.S. 
\rev{Animations of Figures \ref{fig:2}, \ref{appfig:masks}, \ref{appfig:environment}, \ref{appfig:slices}, and \ref{appfig:pyvista} are available in the online Journal
and on FigShare: \dataset[doi:10.6084/m9.figshare.28205066.v1]{https://doi.org/10.6084/m9.figshare.28205066.v1}.
}

\clearpage
\appendix

\section{Robustness of SO brightness peak to continuum subtraction} \label{app:contsub-SO}

\begin{figure*}
\centering
\includegraphics[width=0.9\textwidth]{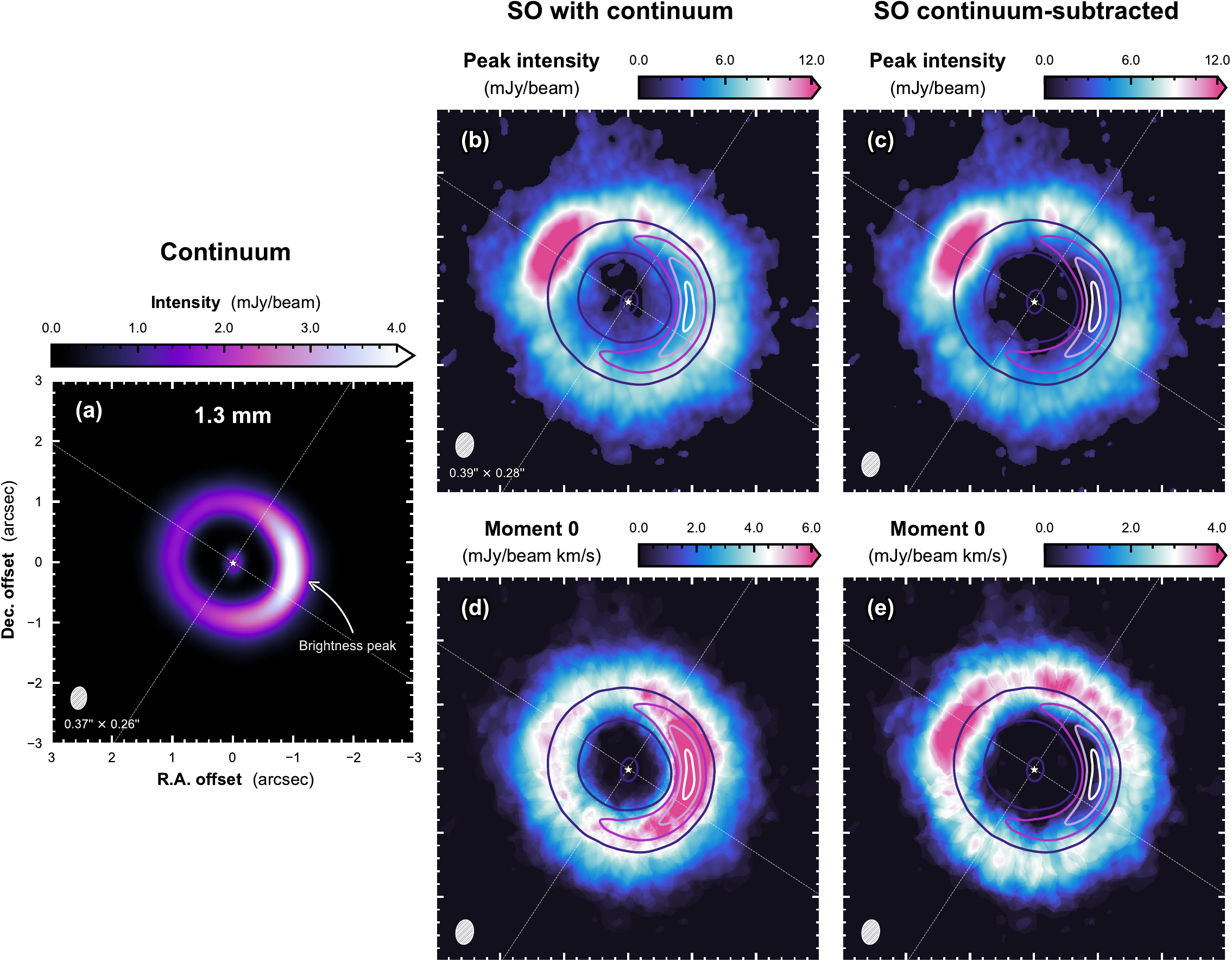}
\caption{\textbf{{The influence of the continuum ring substructure on the observed SO ring substructure in peak intensity vs. integrated intensity maps.}} \textbf{(a)} ALMA continuum image at $233.0$ GHz from the program presented in this work. The continuum is overlaid in contours at $1.0$, $2.0$, $3.0$ and $4.0$ mJy/beam in all other panels.  
\textbf{(b)} ALMA SO $J_{N}=5_6-4_5$ peak intensity map before continuum subtraction, and \textbf{(c)} after continuum subtraction (shown in the main text). 
{These two maps are shown on the same colorbar scale.} 
{\textbf{(d)} ALMA SO $J_{N}=5_6-4_5$ integrated intensity map before and \textbf{(e)} after continuum subtraction.}  
}
\label{appfig:SO-wcont}
\end{figure*}

Figure \ref{appfig:SO-wcont} presents our ALMA continuum image at $1.3$ mm alongside peak intensity {and integrated intensity} maps of SO $J_{N}=5_6-4_5$ imaged with and without continuum subtraction. 
The SO ring peaks exterior to, but still overlaps with, the continuum ring \cite[c.f. Fig. 2 of][]{dutrey2024-abaur-SO}. 
As seen in previous works \cite[e.g.,][]{tang2017-abaur12COspirals, vandermarel2021-diversity-asymmetries, rivieremarichalar2024-rosettastone-3}, the continuum ring exhibits an azimuthal asymmetry, {concentrated on nearly the opposite side of the disk to the brightness peak in SO}.  
We re-imaged the SO data without continuum subtraction to confirm the SO brightness peak is not the gas counterpart of the dust continuum asymmetry. 
{As shown in the top row of Figure \ref{appfig:SO-wcont}, the two \textit{peak} intensity maps are very similar: the inclusion of the continuum extends the ring's apparent width inward, but the exterior substructure is brighter than the continuum and remains unchanged. In contrast, the substructure seen in an \textit{integrated} intensity map of SO is greatly influenced by the presence or subtraction of the continuum, as shown in the bottom row of Figure \ref{appfig:SO-wcont}.}

\section{Keplerian and Anti-Keplerian Masks} \label{app:masks-kep-antikep}

Our process to disentangle the bulk rotating component of $^{12}$CO $J=2-1$ emission from all other emission was inspired by \citet{huang2021-MAPS-GMAur} and broadly follows their approach to isolate the extended $^{12}$CO structures in GM Aur (see their Figure 17 and Appendix C). 

First, we generate a Keplerian mask using the \texttt{keplerian\_mask} script of \citet{teague2020-keplerianmask-zenodo}. The result is a cube of the same dimensions as the data but filled with Boolean values reflecting the {position-position-velocity (PPV)} volume of the $^{12}$CO emitting region \cite[e.g., see \S5.1 of ][]{czekala2021-maps2}. This volume is calculated given a specified stellar mass $M_{\star}$, distance $d$, disk inclination $i$, position angle, systemic velocity $v_{\rm sys}$, outer radius $r_{\rm out}$, constant emission surface slope $z/r$, and intrinsic line width parameters $\Delta V_{0}$ and $\Delta V_{q}$. 
The intrinsic line width profile is described by: 
\begin{equation}
    \Delta V (r) = \Delta V_{\rm 0} \, \Big(\frac{r}{1\arcsec}\Big)^{\Delta V_{q}} \, .
    \label{eqn:kepmask-lineprofile}
\end{equation}
Table \ref{apptab:kepmask-parameters} lists the parameters we adopted. 
Our choices for $r_{\rm out}$, $z/r$, $\Delta V_{\rm 0}$ and $\Delta V_{q}$ were found by experimentation, with the goal of capturing the disk emission at large radii (made tricky by super-Keplerian rotation) while evading non-Keplerian emission at small radii (made tricky by S1). The \texttt{keplerian\_mask} routine additionally offers the option to convolve the mask with a 2D Gaussian beam (of FWHM $\theta_{\rm conv}$) to account for emission broadening, which we opted not to do for simplicity.  

Next, we generate an anti-Keplerian mask as an exact copy of the Keplerian mask but with its Boolean values switched (\texttt{1}s to \texttt{0}s and vice versa). The result is a second cube of the same dimensions as the data, now with a `hole' reflecting the PPV volume of the $^{12}$CO emitting region. 

\begin{figure*}
\centering
\includegraphics[width=0.85\textwidth]{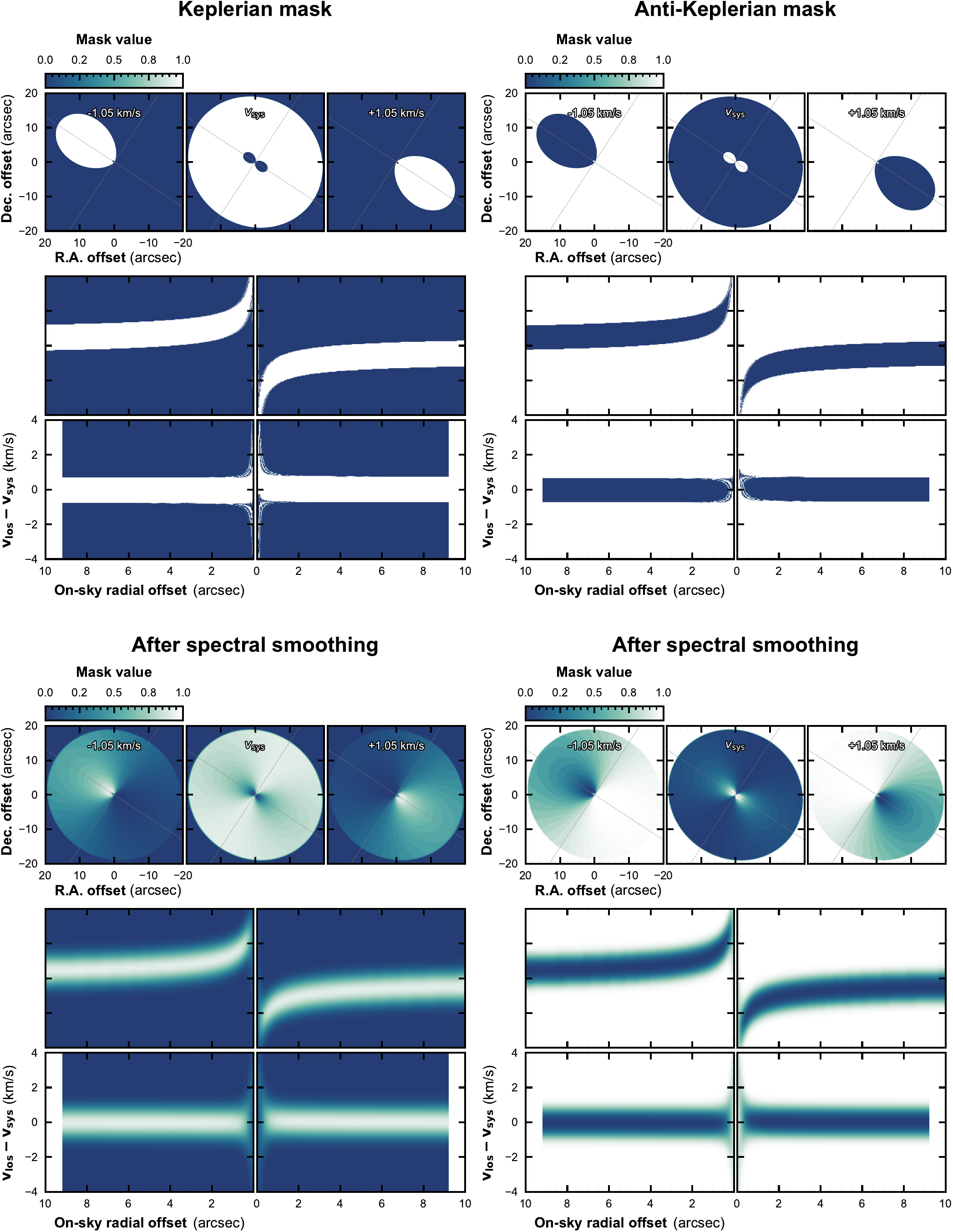}
\caption{\textbf{A presentation of the Keplerian and anti-Keplerian masks.} 
\textbf{Top:} Boolean versions prior to spectral smoothing. 
The topmost panels show three channel maps from the masks at three demonstrative velocities. 
The panels immediately below show PV slices taken through the masks along the disk major and minor axes, 
as in Figure \ref{fig:2}. 
\textbf{Bottom:} Smoothed versions after convolving the spectra in each spatial pixel of the masks with a Gaussian kernel. 
These are used as 3D weighting functions to separate the observed $^{12}$CO emission into its constituent disk and exo-disk components.  
\revv{An animation of the Keplerian and anti-Keplerian masks is \href{https://figshare.com/articles/media/Mapping_the_merging_zone_of_late_infall_in_the_AB_Aur_planet-forming_system/28205066?file=51680891}{available} in the online Journal.
The animation shows the channel maps of the two masks before (top row) and after (bottom row) spectral smoothing, panning through the imaged spectral axis from $-0.198$ to $12.444$ km/s in $0.042$ km/s LSRK velocity channels.}
}
\label{appfig:masks}
\end{figure*}

The top half of Figure \ref{appfig:masks} provides a view of the Keplerian and anti-Keplerian masks at this stage in the process. For what follows, note that the spectrum in each spatial pixel of the Keplerian mask is a top hat function (and in the anti-Keplerian mask it is an upside-down top hat function). 

From here, we perform post-processing on the masks to mitigate artifacts created in the moment maps from the hard (discontinuous) edges of the PPV volumes \cite[c.f. Figure 17 of][]{huang2021-MAPS-GMAur}. 
We spectrally smooth the masks by convolving the spectrum in each spatial pixel with a Gaussian 5 channels wide. This transforms Boolean masks into 3D weighting functions and has the desired effect of softening the edges of the PPV volumes. Insofar as the weights in the edges are modulated (from \texttt{1} to slightly less than \texttt{1}, or from \texttt{0} to slightly more than \texttt{0}), the spectral smoothing introduces `overlap' between the PPV volumes in each cube.  
The bottom half of Figure \ref{appfig:masks} provides a view of the Keplerian and anti-Keplerian masks at the end of the process. 

As a final note of nuance, we considered defining an anti-Keplerian mask as a Keplerian mask of opposite \textit{inclination}; however, the resulting two masks would be identical at the disk minor axis, and ineffective at separating the $^{12}$CO emission components identified here.

\begin{table*}
  \caption{ Keplerian Mask Parameters }
  \label{apptab:kepmask-parameters}
  \begin{tabular}{ccccccccccc}
  \hline
    \textbf{Transition} & $M_{\star}$ $^{a}$ & $d$ $^{b}$ & $i$ $^{c}$ & ${\rm P.A.}$ $^{a}$ & $v_{\rm sys}$ $^{a}$ & $r_{\rm out}$ & $\Delta V_{0}$ & $\Delta V_{q}$ & $z/r$ & $\theta_{\rm conv}$   \\
     & ($M_{\odot}$) & (pc) & ($^{\circ}$) & ($^{\circ}$) & (m/s) & ($\arcsec$) & (m/s) &   &   & ($\arcsec$)   \\ \hline
     $^{12}$CO $J=2-1$ & 2.23 & 155.9 & 23.2 & 236.7 & 5850 & 20.0 & 700.0 & 0.0 & 0.0 & None   \\ \hline
  \end{tabular}
  \par
  \begin{itemize}
    \centering
    \item[$^{a}$] \citet{speedie2024}. $^{b}$ \citet{gaiaDR3-2023}. $^{c}$ \citet{tang2012-abaur-envelope, tang2017-abaur12COspirals}.
  \end{itemize}
\end{table*}

\section{Connection to larger environment} \label{app:environment}

\begin{figure*}
\begin{center}
    \includegraphics[width=0.825\textwidth]{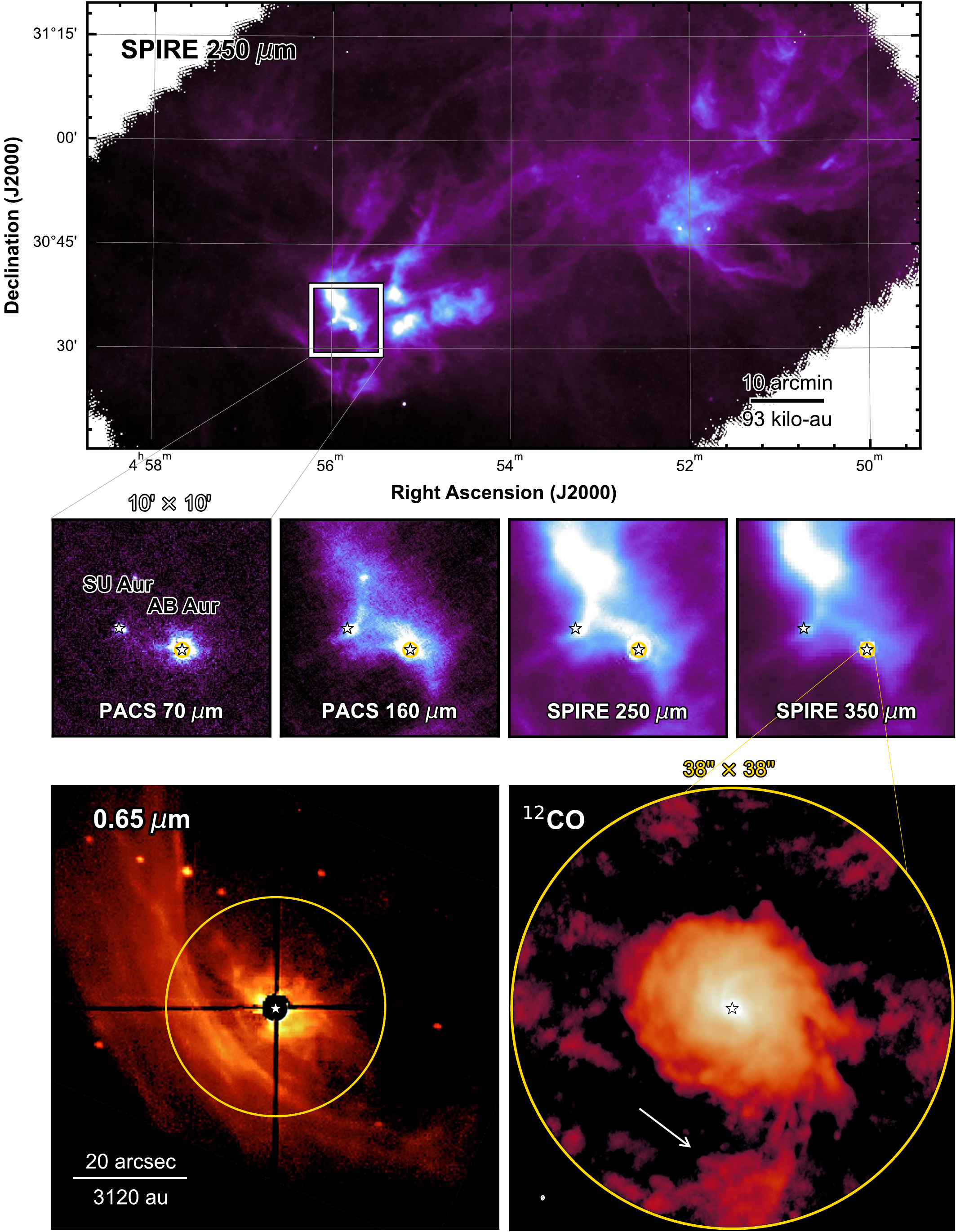}
\end{center}
\caption{\textbf{Multi-wavelength view of AB Aur's kilo-au environment and a tentative tail in $^{12}$CO.} 
\textbf{Top:} Herschel SPIRE $250\,\mu$m map of the L1517 cloud. A scale bar in the bottom right corner shows $10\arcmin$ or $93$ kilo-au. 
\textbf{Middle:} $10\arcmin \times 10\arcmin$ zoom-in toward AB Aur in Herschel PACS and SPIRE maps at $70\,\mu$m, $160\,\mu$m,  $250\,\mu$m and  $350\,\mu$m. 
The location of SU Aurigae is also marked. 
\textbf{Bottom right:} ALMA $^{12}$CO $J=2-1$ moment 0 map, imaged with Briggs \texttt{robust=1.5} and without primary beam correction.  
The white arrow points to the tail. 
\revv{In the online Journal, an animation is \href{https://figshare.com/articles/media/Mapping_the_merging_zone_of_late_infall_in_the_AB_Aur_planet-forming_system/28205066?file=51680894}{available} to show the tail in the $^{12}$CO channel maps. 
In the animation, intensity (horizontal colorbar) spans only  $\pm5\sigma$, and the LSRK velocity of the channel (vertical colorbar) is represented by the background color outside the ALMA FOV.} 
\textbf{Bottom left:} Optical image toward AB Aur at $647$ nm from the University of 
Hawaii 2.2 m Telescope \cite[PI: P. Kalas;][]{grady1999-hst}. 
This image is used with permission of P. Kalas. In all panels,
the yellow circle marks the edge of our imaged ALMA field of view (\S\ref{sec:data}).}
\label{appfig:environment}
\end{figure*}

Figure \ref{appfig:environment} provides a multi-wavelength view of AB Aur's environment from $10^{5}$ au to $10^{2}$ au scales. 
The top panel shows where AB Aur is situated within the L1517 cloud in the Taurus-Auriga complex \citep{luhman2023-taurus-gaia, garufi2024-destinys-taurus, ribas2017-herschel}, in 
photometric observations of dust emission from the Herschel SPIRE instrument \citep{pilbratt2010-herschel, griffin2010-spire} at $250 \, \mu$m. 
The second row of panels zooms into a $10\arcmin \times 10\arcmin$ square, and shows the location of AB Aur in Herschel PACS \citep{poglitsch2010-pacs} and SPIRE maps at increasing wavelengths ($70 \, \mu$m, $160\, \mu$m, $250 \, \mu$m and $350\, \mu$m). 
These four maps are Level 2.5 data products retrieved from the Herschel Science Archive, from the Herschel Key Project Gould Belt Survey \cite[][]{andre2010-KPGT_pandre_1,10.26131/irsa72} with OBSIDs 1342204843 and 1342204844 (OD 492). 
SU Aurigae is within $3\arcmin$ of AB Aur on the sky. 
The bottom row zooms into a $38\arcsec \times 38\arcsec$ square centered on AB Aur, showing the field of view of our ALMA observations outlined in yellow. 
In this panel we present a moment 0 map of $^{12}$CO emission imaged with a Briggs weighting of $\texttt{robust=1.5}$, showing the faint `tail-like' emission structure extending from the disk to the south. We stress that this image has not been primary beam corrected (see below). 
In the bottom left panel, we overlay the ALMA field of view as a yellow circle onto the $R$-band ($647$ nm) image toward AB Aur obtained by the University of Hawaii 2.2 m Telescope \citep{grady1999-hst, nakajima-golimowski1995-palomar} showing the arc-shaped reflection nebula surrounding AB Aur from the east to the south \citep{dullemond2019-cloudlet, kuffmeier2020-late-encounters, gupta2023-reflection-nebulae}.

While a detailed kinematic analysis is outside the scope of this work, we describe that the $^{12}$CO tail is comprised of subcomponents: the first is observed blueshifted of $v_{\rm sys}$ (between $5.010-5.850$ km/s in LSRK) due south of the star, and the second is observed redshifted of $v_{\rm sys}$ (between $5.976-6.942$ km/s in LSRK) ever so slightly counter-clockwise of south. The latter exhibits a substructure like three `whiskers' over several channels. 
{Animated $^{12}$CO channel maps are available (Figure \ref{appfig:environment}) in the online article.}

As discussed in \S\ref{sec:discussion:origins}, our detection of the $^{12}$CO tail suffers from the reduced response of the ALMA 12m antenna at the edge of our imaged FOV, where the sensitivity is $5\times$ lower than at the pointing center, and, as stressed above, the image in Figure \ref{appfig:environment} has not been corrected for this effect. We show the non-primary beam corrected image without the quantification of a colorbar to reflect that it should be considered for morphology only. Aliases of the tail also appear multiple times around the edge of the FOV. Such aliasing artifacts are known to be introduced by bright sources outside the primary beam (ALMA Technical Handbook, \S11.5). 
Observations with a larger FOV and sensitivity to larger spatial scales 
are needed 
to achieve improved sensitivity to extended emission, in order to investigate the $^{12}$CO tail and hypothesized association between the cloudlet.

\section{Radial and Azimuthal PV Slices} \label{app:slices-frames}

\begin{figure*}
\includegraphics[width=\textwidth]{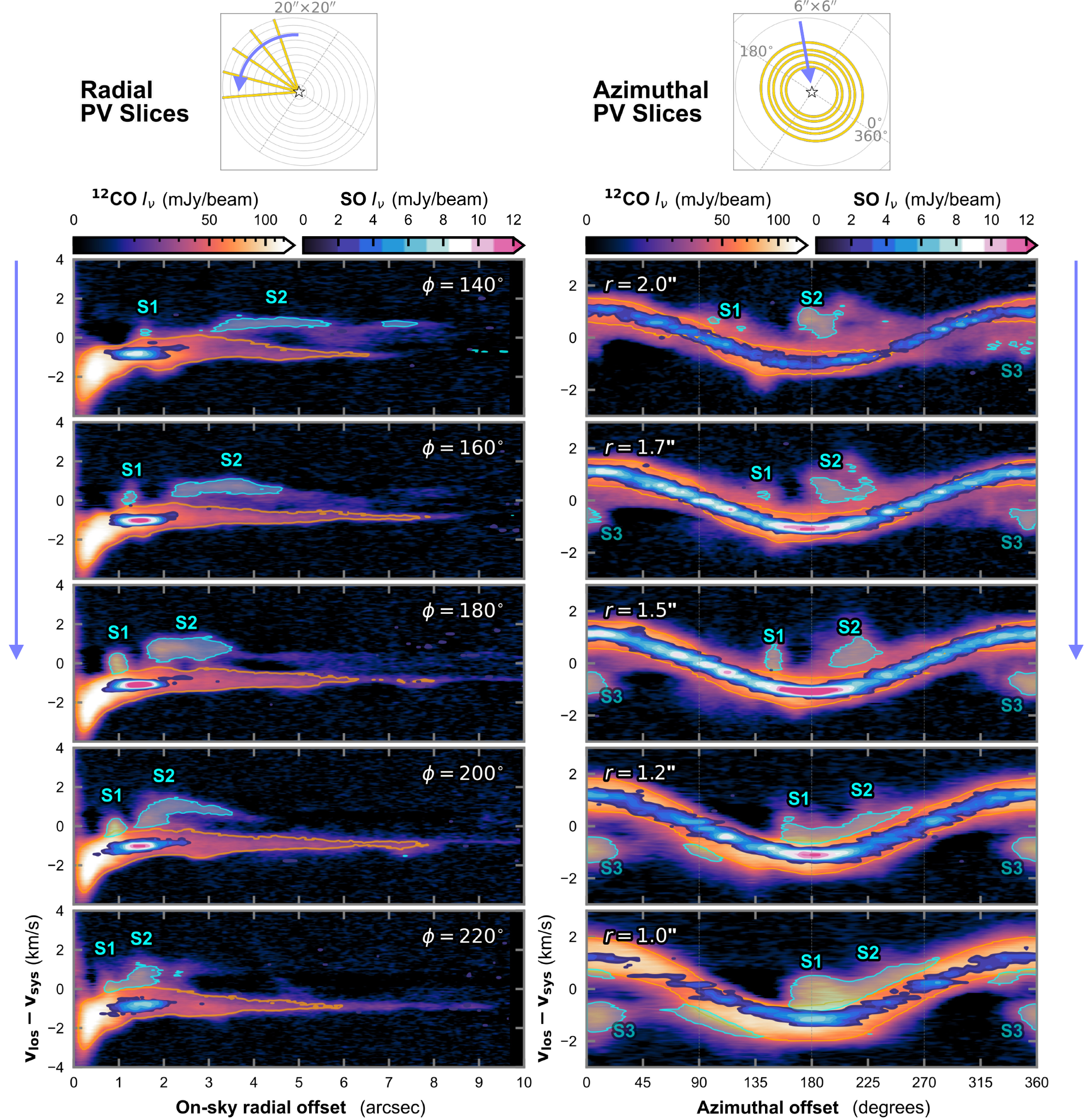}
\caption{\textbf{Radial and azimuthal PV cross sections in $^{12}$CO and SO.} 
The yellow lines and circles in the diagram above each column shows the slice paths of the PV diagrams in the panels below. 
In the left column, panels are ordered top to bottom by their \revv{azimuthal} angle in the direction the disk rotates (\revv{counter-}clockwise). 
In the right column, panels are ordered by decreasing radius. 
PV slices are shown for the three $^{12}$CO cubes: the full unweighted $^{12}$CO cube (colormap), the Keplerian weighted $^{12}$CO cube (orange contours), and the anti-Keplerian weighted $^{12}$CO cube (cyan contours). 
PV slices in the SO $J_{N}=5_6-4_5$ image cube are overplotted as filled contours in increments of $2\sigma$ starting at $3\sigma$. 
\revv{A set of animations for this figure is available in the online Journal. 
Three animations are concatenated and played sequentially. 
At \href{https://figshare.com/articles/media/Mapping_the_merging_zone_of_late_infall_in_the_AB_Aur_planet-forming_system/28205066?file=51680915}{$t=3\mbox{s}$}: The radial PV slices panning through all azimuthal angles (same as Figure \ref{fig:2}).
At \href{https://figshare.com/articles/media/Mapping_the_merging_zone_of_late_infall_in_the_AB_Aur_planet-forming_system/28205066?file=51680909}{$t=29\mbox{s}$}: The azimuthal PV slices, panning from $r=12\arcsec \mbox{ to }0\arcsec$. 
At \href{https://figshare.com/articles/media/Mapping_the_merging_zone_of_late_infall_in_the_AB_Aur_planet-forming_system/28205066?file=51680912}{$t=69\mbox{s}$}: A repeat of the azimuthal PV slices, using a varying colorbar to reveal features at large radii.} 
}
\label{appfig:slices}
\end{figure*}

Figure \ref{appfig:slices} explores the trajectories of S1, S2 and S3 in in radial and azimuthal PV diagrams. 
The left column of Figure \ref{appfig:slices} shows radial PV cross sections at varying azimuths around the eastern (blueshifted) major axis. The right column shows azimuthal PV cross sections at varying radii between $r=1\arcsec-2\arcsec$.  
As in the main text, the background colormap shows slices taken through the (unmasked) $^{12}$CO image cube, and the filled contours are from slices taken through the SO image cube. Orange contours are overplotted using slices through the Keplerian weighted $^{12}$CO cube to delineate emission associated with the disk. 
Cyan contours encircle S1, S2 and S3 from slices through the anti-Keplerian weighted $^{12}$CO cube. 
We inspected a fine and complete grid of PV slice orientations to build an understanding of their $v_{\rm LOS}$ behaviour as a function of azimuth and radius, \rev{available as animations in the online article}. 
A representative selection of slices is shown in Figure \ref{appfig:slices}, demonstrating 
where $^{12}$CO emission from S1 and S2   
becomes indistinguishable from 
the emission from the disk.

\section{Streamline Parameters and 3D Rendering} \label{app:streamline-params}

Table \ref{tab:pims_params} provides the parameters describing the streamline solutions presented in \S\ref{subsubsec:analytic-solutions}.  
The angular frequency of the sphere's rigid-body rotation is represented by $\Omega$, 
and $r_{\rm 0}$ is its radius.   
The initial radial velocity of the `particle' is $v_{r, 0}$, 
while $\phi_{\rm 0}$ and $\theta_{\rm 0}$ are its initial azimuthal and polar angles, respectively. $\theta_{\rm 0}$ is measured from the positive $z$ axis (such that $\theta_{\rm 0}=90^{\circ}$ defines the sphere's {equatorial plane}), and $\phi_{\rm 0}$ is measured from the positive $x$ axis toward the positive $y$ axis.  
The angles $i_{\rm pims}$ and ${\rm PA}_{\rm pims}$ are used in the two rotational transformations to project the streamline's `native' coordinates onto the sky. The inclination $i_{\rm pims}$ defines rotation about the native $x$ axis, and negative inclination corresponds to viewing the sphere's midplane ``from below'', i.e., up from along the negative $z$ axis. 
${\rm PA}_{\rm pims}$ is defined as rotation about the native $y$ axis, where ${\rm PA}_{\rm pims}=0^{\circ}$ aligns minor axis of the midplane with north.

Following the discussion in \S\ref{sec:discussion:SO}, we provide  
the intrinsic velocity components at the terminal ends of the three model streamlines {in the frame of the rigid rotating sphere}: the inward radial velocity $v_{r, \rm term}$, the azimuthal velocity $v_{\phi, \rm term}$, and the downward vertical velocity $v_{z, \rm term}$
(Table \ref{tab:pims_params}). 
The terminal velocities of the S1 and S2 streamlines range $0.7-4.1$ km/s. Table \ref{tab:pims_params} also lists the streamlines' centrifugal radii, $r_{\rm cent}$. 
\citet{sakai2014-nat} connected an observed enhancement of SO to the centrifugal radius in ALMA observations of L1527, which has since been seen in more sources \cite[e.g., ][]{oya2016, sakai2017}.  Our model streamline solution for S1 has a centrifugal radius of $\sim350$ au, conspicuously close to the observed radius of the SO ring ($150-300$ au). For S2 the value is larger ($\sim550$ au). 
We caution direct comparisons here though: since the model is ballistic, angular momentum loss by friction is not considered.

\begin{figure*}
\centering
\includegraphics[width=0.9\textwidth]{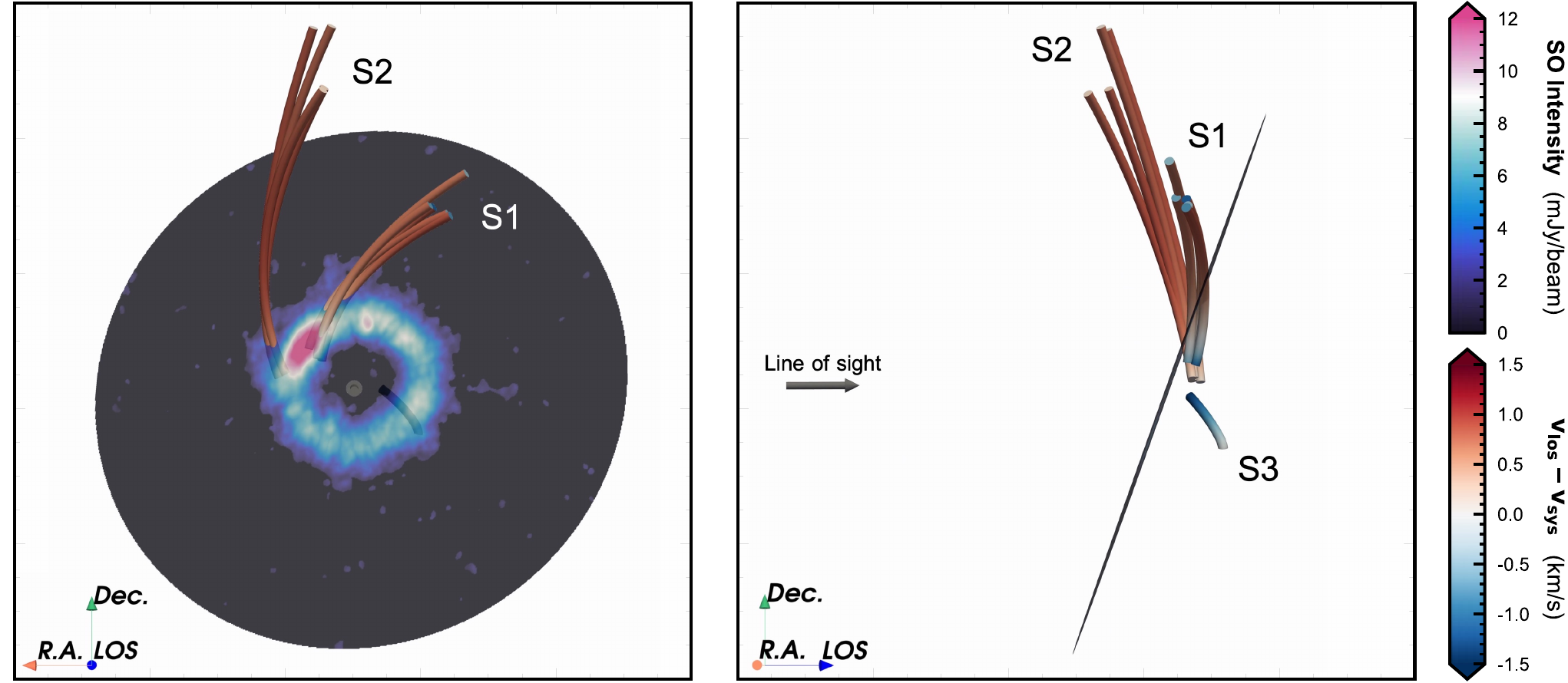}
\caption{\textbf{3D rendering of the AB Aur disk plane and streamline solutions.} 
The left panel shows the system oriented as viewed from Earth (the RA-Dec plane), and the right panel shows a view from within the plane of the sky (the Dec-LOS plane). 
The streamlines are rendered as `tubes' and coloured by their line-of-sight velocity, $v_{\rm LOS}$. 
The SO peak intensity map is rendered onto a disk with same orientation as the AB Aur midplane. The layer is given 80\% transparency to show objects behind the disk. 
\revv{An animation of this figure is \href{https://figshare.com/articles/media/Mapping_the_merging_zone_of_late_infall_in_the_AB_Aur_planet-forming_system/28205066?file=51680903}{available} in the online Journal, in which the viewing angle is rotated $360\degr$ about the declination axis to clearly illustrate the 3D geometry.} 
This rendering was enabled by \texttt{pyvista} \citep{sullivan2019pyvista}. 
}
\label{appfig:pyvista}
\end{figure*}

In Figure~\ref{appfig:pyvista}, we show a rendering of the disk+streamline system in 3D spatial coordinates, $\left\langle {\rm RA}, {\rm Dec}, {\rm LOS} \right\rangle$, 
using the software \texttt{pyvista} \citep{sullivan2019pyvista}. 
The SO peak intensity map is rendered, semi-transparent, onto a planar disk with the orientation of AB Aur's midplane for comparison with the terminal ends of S1 and S2. 
It is significant that despite letting all parameters vary (save for the central mass $M$; \S\ref{subsubsec:pims}), we 
find a family of streamline solutions for each of S1 and S2 that are physically coherent in 3D spatial coordinates.

\begin{center}

\begin{table*}
     \caption{  Parameters describing the PIMS streamline solutions, and resulting properties. }
     \hspace{-6em}
  \begin{tabular}{cccccccc|cccc}
  \hline
 \textbf{Streamline} & $\Omega$    & $r_{\rm 0}$   & $v_{r, 0}$  & $\phi_{\rm 0}$  & $\theta_{\rm 0}$ & $i_{\rm pims}$   & ${\rm PA}_{\rm pims}$ & $r_{\rm cent}$ & $v_{r, \rm term}$ & $v_{\phi, \rm term}$ & $v_{z, \rm term}$   \\
          & (Hz)     & (au) & (km/s) & ($^{\circ}$) & ($^{\circ}$)  & ($^{\circ}$) & ($^{\circ}$) & (au) & (km/s) & (km/s) & (km/s) \\ \hline
S1        & $1.50\times 10^{-11}$ & $650$  & 1.5    & 374   & 80     & -56.8 & 43.3  & 350 & 1.4 & 4.1 & 0.7 \\
S2        & $8.00\times 10^{-12}$ & $1000$ & 1.0    & 30    & 55     & -56.8 & 53.3  & 557 & 1.7 & 3.0 & 2.0 \\
S3        & $1.40\times 10^{-11}$ & $325$  & 0.25   & 290   & 135    & -66.8 & 33.3 & 19 & 5.4 & 1.3 & 3.3 \\
\hline
  \end{tabular} 
    \label{tab:pims_params}
\end{table*}
\end{center}

\clearpage\newpage
\bibliography{new.ms}{}
\bibliographystyle{aasjournal}

\end{document}